\documentclass[11pt,a4paper]{article}
\pdfoutput=1
\usepackage{jheppub}
\usepackage{graphicx}
\usepackage{lipsum}
\usepackage{cancel}
\usepackage{slashed}
\usepackage{soul}
\usepackage{mathtools}
\usepackage{bm}
\usepackage{float}
\usepackage{caption}
\usepackage{amsmath, amssymb}
\usepackage{bbold}
\usepackage[all]{hypcap}
\usepackage{tikz}
\usepackage{tikz-feynman}
\usepackage{pgfplots}
\usepackage{bbm}
\usepackage{xcolor}
\usepackage[font=small,labelfont=bf]{caption}
\usepackage{bbm}
\usepackage{color, colortbl}
\usepackage{subcaption}
\usepackage[font=footnotesize,labelfont=bf]{caption}
\usepackage{array}
\usepackage[export]{adjustbox}

\allowdisplaybreaks

\def\gR{g_{R,3}}
\def\gBL{g_{B-L,3}}
\def\gY{g_{Y,12}}

\def\heavy{{[3]}}
\def\light{{[12]}}

\def\cH{\mathcal{H}}
\def\cG{\mathcal{G}}

\def\modH{h}

\def\PP{\Phi}
\def\hh{H}
\def\ff{F}

\def\fHC{f_{\rm HC}}
\def\LambdaHC{\Lambda_{\rm HC}}

\usepackage{xspace}

\newcolumntype{P}[1]{>{\centering\arraybackslash}p{#1}}
\allowdisplaybreaks
\newcommand{\be}{\begin{equation}}
\newcommand{\ee}{\end{equation}}
\newcommand{\bea}{\begin{eqnarray}}
\newcommand{\eea}{\end{eqnarray}}
\newcommand{\beq}{\begin{equation}}
\newcommand{\eeq}{\end{equation}}
\newcommand{\no}{\nonumber}

\newcommand{\cL}{{\cal L}}

\newcommand{\cO}{{\cal O}}

\begin{document}

\title{
%Composite Higgs with Flavour Deconstruction
Flavour Deconstructing the Composite Higgs
%Alternate title: `Deconstructing the Little Hierarchy Problem' 
}

\author[a]{Sebastiano Covone,}
\author[b]{Joe Davighi,}
\author[a]{Gino Isidori,}
\author[a]{and Marko Pesut}

\affiliation[a]{Physik-Institut, Universit\"at Z\"urich, CH 8057 Z\"urich, Switzerland}
\affiliation[b]{Theoretical Physics Department, CERN, 1211 Geneva, Switzerland}
\emailAdd{sebastiano.covone@physik.uzh.ch}
\emailAdd{joseph.davighi@cern.ch}
\emailAdd{isidori@physik.uzh.ch}
\emailAdd{marko.pesut@physik.uzh.ch}

\abstract{
We present a flavour non-universal extension of the Standard Model combined with the idea of Higgs compositeness. At the TeV scale, the gauge groups $SU(2)_R$ and $U(1)_{B-L}$ are assumed to act in a non-universal manner on light- and third-generation fermions, while the Higgs emerges as a pseudo Nambu-Goldstone boson of the spontaneous global symmetry breaking  $Sp(4)\to SU(2)_L\times SU(2)_R^{[3]}$, attributed to new strong dynamics. 
The flavour deconstruction means the couplings of the light families to the composite sector (and therefore the pNGB Higgs) are suppressed by powers of a heavy mass scale (from which the Higgs is nevertheless shielded by compositeness), explaining the flavour puzzle.
We present a detailed analysis of the radiatively generated Higgs potential, showing how this intrinsically-flavoured framework has the ingredients to justify the unavoidable tuning in the Higgs potential necessary to separate electroweak and composite scales. This happens for large enough values of the $SU(2)_R^{[3]}$ gauge coupling and light enough flavoured gauge bosons resulting from the deconstruction, whose phenomenology is also investigated. The model is compatible with current experimental bounds and predicts new states at the TeV scale, which are within the reach of near future experimental searches. 
}

\begin{flushright}
 CERN-TH-2024-112
 \end{flushright}

\maketitle
\section{Introduction}

The Higgs sector of the Standard Model (SM) faces two significant structural problems. The first one, the {\em flavour puzzle}, is the unexplained hierarchical structure of the Yukawa couplings, which span five orders of magnitude. To address this issue, as well as other deficits of the SM,
it is natural to posit the existence of new heavy dynamics. This assumption, in turn, inevitably leads to the second structural problem of the Higgs sector: the {\em electroweak hierarchy problem} or the instability of the quadratic term in the Higgs potential caused by the presence of heavy degrees of freedom.

Traditional model-building approaches tend to separate the solution to these two problems. This is achieved assuming flavour-blind dynamics close to the electroweak  (EW) scale 
to screen the Higgs sector from heavy dynamics, and flavour-full dynamics at much higher energies to generate the flavour hierarchies. From a general quantum field theory perspective, this is certainly a viable option: the Yukawa couplings are marginal operators whose origin could be attributed to very high energy scales. This separation leads to the so-called Minimal Flavour Violating (MFV) paradigm~\cite{DAmbrosio:2002vsn}, which has the important phenomenological advantage of minimising corrections to the precisely tested sector of rare flavour-changing neutral-current processes (FCNCs). However, the advantage of MFV in the pre-LHC era has turned into a disadvantage nowadays~\cite{Davighi:2023iks,Allwicher:2023shc}.
%, given the strong direct-searches bounds on New Physics (NP) coupled universally to all fermion families.
%make it less effective for solving the EW hierarchy problem.}} The high-energy searches at the LHC \gino{seem to indicate that if we are interested in minimising the fine-tuning in the Higgs sector, hence we insist on TeV-scale NP coupled to the Higgs sector, then is better to give up the assumption of flavour universality and assume that, beside the Higgs,  NP couples mainly to third-generation fermions. This, in turn, suggests the flavour puzzle and electroweak hierarchy problem are inexorably related and must be addressed simultaneously ~\cite{Davighi:2023iks}.} \mar{I would rephrase a bit: High-energy searches at the LHC suggest that if we aim to minimize fine-tuning in the Higgs sector—insisting on TeV-scale NP coupled to the Higgs—it is advantageous to give up the assumption of flavour universality. Instead, one should assume that, in addition to the Higgs, NP primarily couples to third-generation fermions. This, in turn, suggests the flavour puzzle and electroweak hierarchy problem are inexorably related and must be addressed simultaneously ~\cite{Davighi:2023iks}.} 
%%
The null results coming from high-energy searches 
at the LHC
suggest that if we aim to minimize fine-tuning in the Higgs sector -- which requires new physics (NP) close by the TeV scale coupled to the Higgs field -- then it is advantageous to give up the assumption of flavour universality: NP coupled primarily to third-generation fermions and the Higgs can more easily escape the constraints from direct search and thus be lighter and more natural.
%Instead, if NP couples primarily to third generation fermions and the Higgs then it can more easily escape direct-searches bounds, hence be lighter and so more natural.
This instrinsic flavour non-universality, in turn, suggests the flavour puzzle and electroweak hierarchy problem could be inexorably related, and motivates the consideration of NP models that address both simultaneously~\cite{Davighi:2023iks}. 
In this paper, we combine the hypothesis of a composite Higgs, in the spirit of the minimal composite Higgs model originally proposed in~\cite{Contino:2003ve,Agashe:2004rs},  with the idea of flavour deconstruction, along the lines developed recently in~\cite{Davighi:2023iks,Barbieri:2023qpf}. As we will demonstrate, combining these two hypotheses provides significant benefits, as it allows new-physics coupled dominantly to the third generation, responsible for the stability of the Higgs sector, to lie in the few TeV domain.

In composite models the Higgs is assumed to be the pseudo Nambu-Goldstone boson (pNGB) of a new strongly interacting sector, much like the pions in QCD (see~\cite{Panico:2015jxa} for a comprehensive review).  Being a  composite state, the Higgs is protected from quantum corrections induced by dynamics occurring at energies above the compositeness scale.  Moreover, the shift symmetry associated to Goldstone bosons allows a scale separation between the Higgs mass term and the confinement scale of the new strong dynamics~\cite{Giudice:2007fh}.
This scale separation cannot be large, and the Higgs mass term (and the electroweak scale) is radiatively generated by the unavoidable breaking of the global symmetry of the composite sector induced by the Yukawa couplings and coupling to the SM gauge fields. Present constraints from EW observables and Higgs physics, as well as the absence of direct signals of new dynamics, imply that the scale associated to the new dynamics must lie above about~1~TeV. This {\em little hierarchy} between the EW and the composite scales would still be acceptable to minimise fine tuning in the Higgs potential. 

However, a compositeness scale in the TeV range is ruled out by FCNC data if the flavour structure of the model is generic. In this class of models, the SM fermions are the result of a linear mixing between elementary fields and composite states (partial compositeness) that delivers the effective Yukawa couplings. An extensive and updated phenomenological analysis of this mechanism has been presented in~\cite{Glioti:2024hye}, where it has been show that global flavour symmetries are a key ingredient to allow a small scale separation between the elementary and composite dynamics. Not surprisingly, the most efficient way to achieve this goal is via global $U(2)^n$ symmetries acting on the light fermions only, as originally proposed in~\cite{Barbieri:2011ci,Barbieri:2012uh,Isidori:2012ts,Redi:2012uj}. This is where the mechanism of flavour deconstruction becomes relevant, as it provides a natural origin for these otherwise {\em ad hoc} flavour symmetries. Namely, they emerge as an accidental low-energy property of a manifestly flavour non-universal gauge group in the ultraviolet (UV). 

The hypothesis of flavour non-universal gauge interactions as the origin of the flavour hierarchies is quite old and has been pursued in different contexts (see {\em e.g.}~\cite{Li:1981nk,Dvali:2000ha,Craig:2011yk,Panico:2016ull}). The last few years have witnessed a renewed phenomenological interest in this approach~\cite{Bordone:2017bld,Greljo:2018tuh,Fuentes-Martin:2020pww,Fuentes-Martin:2020bnh,Davighi:2022fer,Fuentes-Martin:2022xnb,FernandezNavarro:2022gst,FernandezNavarro:2023rhv,Davighi:2022bqf,Greljo:2024ovt,Capdevila:2024gki,Fuentes-Martin:2024fpx} largely motivated by the the so-called $B$-anomalies. While the significance of these deviations from the SM remains the subject of ongoing debate (see~\cite{Koppenburg:2023ndc} for a recent update), the corresponding model-building activity has provided interesting explicit examples of models featuring TeV-scale dynamics, coupled mainly to the third generation and fully compatible with present observations, that can be motivated to address the flavour puzzle.  

The general ingredient of flavour deconstruction is the breaking of a flavour non-universal gauge group of the type $G_{A} \times G_{B}$ into its flavour-diagonal and maximal subgroup $G_{A+B}$.\footnote{That the flavour-diagonal subgroup is left unbroken is completely generic when $G$ is semi-simple~\cite{Craig:2017cda}, since a lemma of Goursat~\cite{Goursat1889,bauer2015generalized} implies there are no other non-trivial subgroups isomorphic to $G$.}
If $G_{A}$ acts only on the light families (and $G_B$ on the third one) we naturally achieve the accidental $U(2)^n$ symmetries acting on the light fermion families, which are the key ingredient to minimise the tight bounds on new physics dictated by FCNCs.  
The class of deconstructed model we are interested in in this paper was identified in~\cite{Davighi:2023iks} as being an especially natural candidate, and further developed in~\cite{Barbieri:2023qpf}. It is based on the deconstruction of $SU(2)_R$ and $U(1)_{B-L}$, where $SU(2)_L$ is kept flavour-universal. This deconstruction pattern leads to a parametric structure for the Yukawa couplings of the type 
    \begin{equation} \label{eq:yukawa-texture}
        Y_{u,d,e} \sim \begin{pmatrix}
            \epsilon_R & \epsilon_L \\
            \epsilon_R \epsilon_L & 1
        \end{pmatrix}
    \end{equation}
where, for simplicity, we do not distinguish first and second generations. The parameters $\epsilon_{L,R}$ appearing in (\ref{eq:yukawa-texture}) are ratios of the vacuum expectation values (VEVs)  of the scalar (link) fields responsible for the spontaneous symmetry breaking $G_{A} \times G_{B} \to G_{A+B}$, over the mass of appropriate heavy fermions. The presumed smallness of these ratios is the origin of the flavour hierarchies. If one considered only flavour hierarchies and ignored the $B$-anomalies, nothing anchors the overall scale of these new degrees of freedom. However, the situation changes if we aim at addressing also the electroweak hierarchy problem.  As advocated in~\cite{Allwicher:2020esa,Davighi:2023iks,Davighi:2023evx,Davighi:2023xqn}, the overall scale of the deconstructed gauge dynamics should be anchored by its impact on the Higgs sector, which is unavoidable since (at least some) the link fields are charged under the EW symmetry group, and hence couple directly to the Higgs.
While this argument was put forward in~\cite{Allwicher:2020esa,Davighi:2023iks,Davighi:2023evx,Davighi:2023xqn} using only semi-quantitative finite-naturalness arguments, in this paper we provide a quantitative concrete analysis implementing flavour deconstruction in the context of the minimal composite Higgs model.

Given the deconstruction pattern sketched above, where the Higgs and third-generation fermions are charged under the family-specific $SU(2)_R^{[3]}$ symmetry, the minimal embedding of the Higgs as a pNGB is realised assuming a global $Sp(4)$ symmetry in the composite sector, spontaneously broken to $SU(2)_L \times SU(2)_R^{[3]}$. While many features of the flavour-universal framework go through unchanged, two important differences arise.
First, the freedom in choosing the value of the $SU(2)_R^{[3]}$ gauge coupling\footnote{We choose to gauge the full $SU(2)_R^\heavy$ symmetry here, but one could just as appealingly consider a version in which only $U(1)_R^\heavy \subset SU(2)_R^\heavy$ is gauged. The former is motivated by a desire to move towards semi-simple gauge groups in the UV and absorb $U(1)$ factors; on the other hand, the latter can account for the top-bottom mass splitting without fine-tuning (see sect.~\ref{sect:topY}). }
provides a new ingredient to achieve the little hierarchy ({\em i.e.}~a small 
ratio between the EW scale and the composite one). This is obtained via an accidental cancellation between fermion and gauge contributions in the Higgs potential, which are predicted to have opposite sign. Second, this same cancellation requires the masses of the heavy gauge bosons resulting from deconstruction (in particular the $W^\pm_R$ and $Z_R$ bosons)  to be in the few TeV range, well below any of the composite states except for the Higgs and the top partners. 
As we shall see, this expectation is fully consistent with present data for large enough values of the $SU(2)_R^{[3]}$ coupling, which is also what the tuning in the potential asks for. Merging partial compositeness and flavour deconstruction, we achieve a framework that is more predictive than when considering the two hypotheses separately and is fully compatible with present observations.

This work is not the first attempt to merge the idea of a composite Higgs sector and flavour non-universal gauge interactions. Interesting alternative proposals, based on different non-universal gauge groups and/or different symmetries of the strong sector, have been presented in~\cite{Fuentes-Martin:2020bnh,Fuentes-Martin:2022xnb,Chung:2021ekz,Chung:2021fpc,Bally:2022naz,Chung:2023gcm,Chung:2023iwj}. Our choice, which is dictated by minimality on both fronts ({\em i.e.}~both local and global symmetries), is particularly well-suited to explore analytically the interplay of the two hypotheses in the generation of the Higgs potential. We are indeed able to compute the latter in a simple analytic form, with a minimal set of assumptions about the strong dynamics.

The paper is organised as follows. In Section \ref{sec::definitionmodel}, we introduce the UV gauge group and the field content of the model; we describe the two-step symmetry-breaking pattern and the main features of the construction. Section \ref{sec:composite} provides a detailed analysis of the composite sector: we introduce the formalism to describe the pNGB dynamics, the mechanism of partial compositeness, and present a detailed analysis of the different contributions to the Higgs potential. Finally, the phenomenological implications of the model are present in Section \ref{sec:pheno}, with particular focus on the effects generated by the massive gauge bosons resulting from flavour deconstruction, which are the distinctive feature of this setup. A summary of the main findings and an outlook on future directions are presented in Section~\ref{sec::conclusion}.

\section{Definition and main features of the model \label{sec::definitionmodel}}

In this section we introduce the main ingredients of the model  and summarise its 
main features. We start by introducing the field content and gauge group, and then summarise the chain of symmetry-breaking steps occurring at different energy scales.

\subsection{Field content and gauge group}

As anticipated, we assume a strong sector that gives rise to the minimal global symmetry breaking pattern to deliver a pNGB Higgs, namely
\be \label{ssbcomposite}
\cG \equiv Sp(4) \xrightarrow{\Lambda_{\rm HC}} SU(2)_L \times SU(2)_R^{\heavy} \equiv \cH  \, .
\ee
We assume the entire group $\cH$ to be gauged. Here and in the following, the 
upper index on a gauge group factor (as in $SU(2)_R^\heavy$)  denotes a flavour non-universal gauge symmetry acting only on the given family of chiral SM-like fermions. 
The spontaneous symmetry breaking (SSB) breaking (\ref{ssbcomposite}) occurs at a scale 
\begin{equation}
\Lambda_{\rm HC} \equiv 4\pi \fHC \gg v.
\end{equation}
 We envisage some strongly coupled gauge sector that triggers this transition dynamically, with some {\em hypercolour} gauge group $G_{\text{HC}}$ (that we do not specify)  underlying the global symmetry $\cG$ in the UV.\footnote{
As is well-known, this minimal $\mathfrak{so}(5) \to \mathfrak{so}(4)$ breaking pattern does not arise straightforwardly from a chiral condensate in a QCD-like hypercolour theory -- in contrast to less minimal options such as the $\mathfrak{su}(4) \to \mathfrak{sp}(4)$ breaking pattern that delivers the Higgs plus an extra gauge singlet pNGB~\cite{Gripaios:2009pe}, which can naturally emerge from an $Sp$ gauge theory. Nonetheless with a little more model-building, the minimal option can be realised from a fundamental gauge theory. For example, by including explicit $\mathfrak{su}(4)$ breaking interactions in the $\mathfrak{su}(4)/ \mathfrak{sp}(4)$ model, one can lift the extra singlet and give an effective description with the minimal coset structure, as proposed in~\cite{Setford:2017csx} (see also~\cite{Davighi:2018xwn}). }
Note that, at the Lie algebra level, we have $\mathfrak{sp}(4) \cong \mathfrak{so}(5)$ and $\mathfrak{su}(2)\oplus \mathfrak{su}(2)\cong \mathfrak{so}(4)$, so this is usually referred to as the $SO(5)/SO(4)$ breaking pattern, which is a slight abuse of notation. 
Interestingly, we do not have the option of imposing an additional semi-direct product between $\mathcal{H}$ and $\mathbb{Z}_2$ (that acts by exchanging the two $SU(2)$ factors~\cite{Gripaios:2014pqa}), simply because that would be inconsistent with the representation of the scalar link field $\Sigma_R$ needed for the flavour deconstruction.\footnote{Such a $\mathbb{Z}_2$ exchange symmetry was considered desirable in the pre-LHC era of composite Higgs model building, to evade the LEP-II $Z\to b\bar{b}$ constraints that are strong when the scale separation $v/\fHC$ (and hence the requisite tuning) is small~\cite{Agashe:2006at}, as was then viable. Now that $\fHC$ is constrained to be at least the TeV scale by LHC data, the LEP-II constraints from $Z\to b \bar{b}$ do not play as big a role.}

Including also the part of the gauge symmetry acting on the light families, the relevant symmetry group above the scale $\Lambda_{\rm HC}$  has the form
$G_{\text{HC}} \times G_{\text{elem}}$,
where 
\be 
\label{eq:full-symmetry}
G_{\text{elem}} = SU(3)_c \times  SU(2)_L \times SU(2)_R^{\heavy}  \times U(1)_{B-L}^\heavy \times U(1)_Y^\light  \,.
\ee 
The flavour deconstruction of $B-L$ is, as we shall see, motivated by the SM flavour puzzle.
 
The matter content of the model can be decomposed into two main sectors: the elementary and the composite one. 
The elementary sector describes fundamental fields which are not charged under $G_{\text{HC}}$
and are well defined also above the scale $\Lambda_{\rm HC}$. Its structure is composed by (see Table~\ref{tab:Matter_Content}):
\begin{itemize}
\item {\bf Light-family chiral fermions:} The first and second generation   fermions are charged under  $SU(2)_L \times U(1)_Y^\light$: the charges are exactly as in the SM case, but for the replacement of the universal hypercharge with $U(1)_Y^\light$.
The $SU(2)_L$ doublets have a 
mass mixing with appropriate vector-like fermions of the composite sector.
\item {\bf Third-family chiral fermions:} 
These fields are charged under  $SU(2)_L \times SU(2)_R^\heavy 
\times U(1)_{B-L}^\heavy$: they consist of  two $SU(2)_L$ doublets ($q_L$ and $\ell_L$)
and two $SU(2)_R^\heavy$ doublets ($q_R$ and $\ell_R$), all equally 
charged under $U(1)_{B-L}^\heavy$. They all have a mass mixing with appropriate vector-like fermions of the composite sector. 
\item {\bf Third-family vector-like fermions:} For each third-family chiral fermion an elementary 
vector-like fermion with the same transformation properties under 
$SU(2)_L \times SU(2)_R^\heavy$
is introduced. When integrated out, these fields generate the suppressed entries of the Yukawa couplings involving light 
families, via appropriate higher-dimensional operators.
\item {\bf Scalar link fields:} The breaking of the non-universal gauge group to the universal SM group occurs via the scalar fields $\Sigma_R$ and $\Omega_{q,\ell}$, with transformation properties as shown
in Table~\ref{tab:Matter_Content}. 
\end{itemize}

\begin{table}[t]
\begin{center}{\small
\begin{tabular}{|c|c||cc|cc|}
\hline
 \multicolumn{2}{|c||}{Elementary fields} &  $U(1)_{B-L}^\heavy$ & $U(1)_Y^\light$ & $SU(2)_L$ & $SU(2)_R^\heavy$ \\
\hline
chiral  & $q_L^\light$  & $0$ & $1/6$ & $\bf{2}$ & $\bf{1}$  \\
light quarks 
        & $u_R^\light$  & $0$ & $2/3$ & $\bf{1}$ & $\bf{1}$  \\ 
        & $d_R^\light$  & $0$ & $-1/3$ & $\bf{1}$ & $\bf{1}$  \\ 
\hline
chiral  & $q_L^\heavy$  & $1/6$ & $0$ & $\bf{2}$ & $\bf{1}$  \\
3$^{\rm rd}$ gen. quarks
        & $q_R^\heavy$  & $1/6$ & $0$ & $\bf{1}$ & $\bf{2}$  \\ \hline\hline
vector-like     
    & $F^q_L$  & $1/6$ & $0$ & $\bf{2}$ & $\bf{1}$  \\
quarks    
    & $F^q_R$  & $0$ & $1/6$ & $\bf{1}$ & $\bf{2}$  \\
\hline \hline
scalar & $\Sigma_R$  & $0$ & $ 1/2$ & $\bf{1}$ & $\bf{2}$ \\
link fields & $\Omega_q$  & $-1/6$ & $1/6$& $\bf{1}$ & $\bf{1}$ \\
 & $\Omega_\ell$  & $1/2$ & $-1/2$ & $\bf{1}$ & $\bf{1}$ \\
\hline
\end{tabular} 
}
\end{center}
\caption{Matter content of the elementary sector. For simplicity, among the fermions 
only the quarks are shown. Note our non-standard choice of normalisation for 
$B$ and $L$ charges in the $U(1)_{B-L}^\heavy$ group,
which are chosen to coincide with their appearance in the SM hypercharge.
The final link field $\Omega_\ell$ is not strictly needed in the model; its presence (motivated perhaps by a desire for quark-lepton unification) means the charged lepton Yukawa parametrically mirrors the quark Yukawa matrices, namely matching (\ref{eq:yukawa-texture}).
\label{tab:Matter_Content}}

\end{table}

\noindent
The composite sector is well defined only below the scale $\Lambda_{\rm HC}$: it describes bound states of fields charged under the hypercolour gauge group which, by construction, form complete representations of $\cH$.  
Among them, the states playing a key role in the construction are:
\begin{itemize}
\item {\bf The Higgs field}, namely the pNGB of the global symmetry breaking $\cG \to \cH$,  that will itself trigger EW symmetry breaking upon acquiring its own vacuum expectation value.
\item {\bf The lightest vector resonances},
namely the lightest composite   
spin-1 states transforming under $\cH$ as the corresponding gauge bosons. 
Their mass, $M_\rho$, is assumed to be  below  the scale $\LambdaHC$.

\item {\bf The top partners},
namely composite spin-$\frac{1}{2}$ states 
 with mass $M_T = O(1) \times \fHC$ that, when integrated out, 
provide the largest contribution to the effective top-quark Yukawa coupling. 
\end{itemize}

\subsection{Symmetry-breaking pattern}

The breaking to the flavour universal electroweak gauge group is realised by the VEVs of the elementary scalar link fields $\Sigma_R$  and 
$\Omega_{q,\ell}$:
\begin{align}
\label{eq:linkVEVs1}
&SU(2)_R^\heavy \times U(1)_{T_R^3}^\light \xrightarrow{\langle \Sigma_R \rangle} U(1)_{T_R^\heavy} \\
&U(1)_{B-L}^\heavy \times U(1)_{B-L}^\light \xrightarrow{\langle \Omega_{q,\ell} \rangle} U(1)_{B-L} \, .
\label{eq:linkVEVs2}
\end{align}
Recall that (third-family) hypercharge is given by $Y^\heavy=T_{R^3}^\heavy  + (B-L)^\heavy$. This symmetry-breaking pattern leads to 4 massive gauge bosons: two $Z^\prime$, 
associated to the neutral $B-L$ and $T^3_R$ generators, and a pair of $W^\pm_R$. The masses and couplings of these heavy gauge bosons will be given in \S \ref{sec:pheno} when we discuss phenomenology.

\begin{figure}[t]
\begin{center}
\begin{tabular}{c|c|c}
Energy & Linearly-realised Symmetry &  Effective interactions \\ \hline
&   &   \\
$100$~TeV &   & $\qquad\qquad\qquad\qquad\qquad\qquad\qquad$    \\ 
&  \small{ 
 $\textcolor{blue}{Sp(4)} \supset \textcolor{red}{
SU(2)_L\times SU(2)_R^\heavy}$ }  &   \\[2pt]
& \small{ $U(1)_{B-L}^\heavy \times  U(1)_Y^\light$  }  &   \\[15pt]
$10$~TeV &   &   \\[15pt] 
&  \small{ 
 $\textcolor{blue}{SU(2)_L \times SU(2)_R} = \textcolor{red}{
SU(2)_L\times SU(2)_R^\heavy}$ }  \\[2pt]
& \small{ $U(1)_{B-L}^\heavy \times  U(1)_Y^\light$  }  &   \\[15pt]
$1$~TeV &   &   \\[25pt] 
&  {\small $SU(2)_L\times U(1)_Y$ }  &   \\[15pt]
$100$~GeV &    &   \\[15pt] 
\end{tabular}
\end{center}
\vskip -8.5 cm
\includegraphics[width=0.38\textwidth, right]{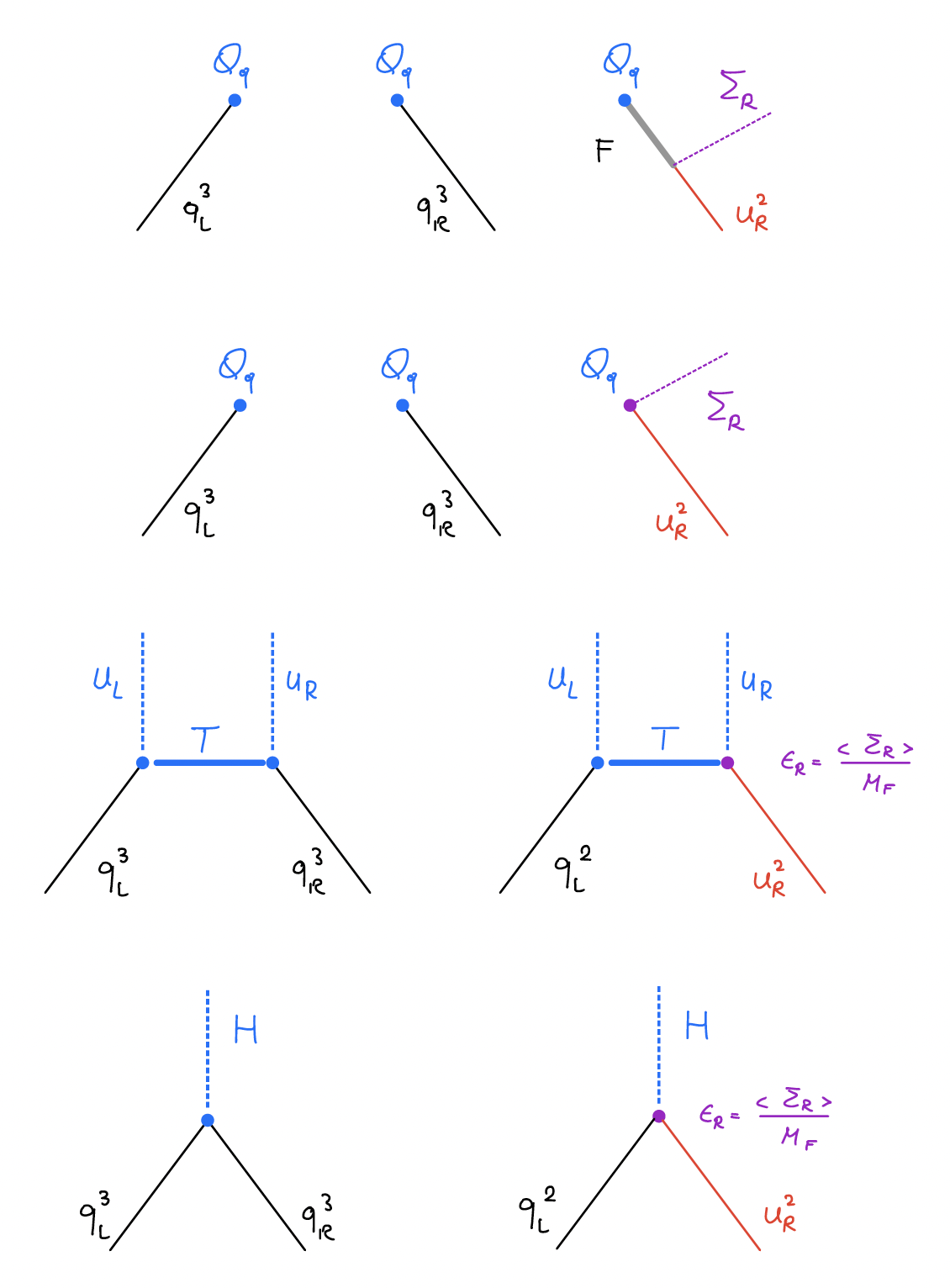}
\vskip 0.5 cm
    \caption{Schematic representation of the 
    symmetry breaking chain.  In the second column, the terms in blue denote   
     global symmetries of the strong sector, while the corresponding gauged subgroups are indicated in red.  
     The interaction terms on the right are those contributing to third- and second-family quark Yukawa couplings ($u_{L,R}$ denote Goldstone boson fields, see \S \ref{sec:composite}).
    \label{fig:SSB1}}
\end{figure}

In parallel to the symmetry breaking induced by the link fields in Eqs.~(\ref{eq:linkVEVs1})--(\ref{eq:linkVEVs2}), the global symmetry breaking $\cG \to \cH$ occurring in the hypercolor sector provides masses to all the composite states but for the pNGB ({\em i.e.}~the SM-like Higgs). As we shall see, the massive $Z^\prime$ and the $W^\pm_R$ are predicted to be  lighter than all the massive composite states. 

The symmetry breaking chain and the  degrees of freedom relevant at different energy scales, 
illustrated in Fig.~\ref{fig:SSB1}, can be summarised  in four main steps as listed below.
\begin{itemize}
\item[I.]
The flavour structure is seeded at high energies ($\gtrsim 100$ TeV or so) by interactions involving the elementary chiral fermions and appropriate fermionic operators 
in the strong sector, denoted $\cO_f$ ($f=q,\ell$). We remain agnostic
on the latter except for their transformation properties 
under $G_{\text{elem}}$.
The non-universal gauge symmetry dictates that only third-generation chiral and 
VL fermions have a
non-suppressed linear mixing with 
the $\cO_f$. A key observation at this stage is that appropriate products of VL fermions and link fields have the same gauge transformation properties as the light chiral fermions.
In particular, in the quark sector we have 
\be
F^q_R \times \Sigma_R \sim 
     u_R^\light\,, 
\qquad 
F^q_R \times \Sigma^c_R \sim 
     d_R^\light\,, 
\qquad      
F^q_L \times \Omega_q \sim q_L^\light\,.
\ee
where $\Sigma_R^c = i \sigma_2 \Sigma_R^*$.
This allow us to generate a non-local coupling of the light chiral fermions to 
 the $\cO_f$ via the exchange of VL fermions (see Fig.~\ref{fig:SSB1}).

\item[II.] Below about 100 TeV, and above the confinement scale $\Lambda_{\rm HC}$, the elementary VL fermions can be integrated out. This leads to the appearence of higher-dimensional operators involving the light chiral fermions and the strong sector.  When, at lower scales, the link fields acquire a VEV, these higher-dimensional operators result in the effective suppression factors
\be
\epsilon_R = \frac{ v_R }{M_F}
\equiv
\frac{\sqrt{2} |\langle \Sigma_R \rangle| }{M_F}
\,,
\qquad 
\epsilon_L = 
\frac{ v_\Omega }{M_F} 
\equiv
\frac{\sqrt{2} \langle \Omega_{q} \rangle}{M_F}
\,.
\ee
Setting $\epsilon_R= O(m_c/m_t) = O(10^{-2})$ and $\epsilon_L= O(|V_{cb}|) = O(10^{-1})$
we obtain the ingredients to achieve, naturally, the hierarchical structure of the Yukawa couplings.

\item[III.] Below $\LambdaHC = O(10)$~TeV and above 
$\fHC = O(1)$~TeV the strong sector leads  to the global $\cG \to \cH$ breaking that deliver  the SM-like Higgs filed as pNGB. 
At the same time,  the two point function $\langle 0| T\{\bar\cO_f(x) \cO_f(0)\} | 0\rangle$ is assumed to be dominated (at low momentum transfer) by the exchange of 
a few light spin-1/2 composite states. For the operators coupled to  third-generation quarks these composite states are the so-called top partners.   
\item[IV.]
Around and below the scale $\fHC$,
having integrated out the top partners and 
frozen the link fields to their VEVs,
we end up with an effective SM-like Yukawa interaction and an effective Higgs
potential. The latter is radiatively generated by couplings/dynamics that explicit
break $\cG$, namely: i) the mass-mixing in the fermion sector, ii)
the gauging of $\cH$, iii) the VEV of $\Sigma_R$. 
\end{itemize}

The first two steps related to the flavour hierarchies are conceptually very similar to what was discussed in refs.~\cite{Davighi:2023iks,Barbieri:2023qpf}, in the context of an elementary Higgs sector. The main difference is that the couplings of the fermions to the Higgs are  replaced  by  couplings to the strong sector.
This has interesting implications for the Higgs hierarchy problem: since above $\LambdaHC$ there is no Higgs field, the heavy elementary VL fermions do not destabilise the Higgs mass term.  In other words, the merging of composite dynamics with flavour deconstruction removes any quadratic sensitivity of $m_H^2$ to the highest mass scale ($M_F \sim 100$ TeV) in this theory of flavour (as might otherwise be na\"ively estimated in a theory with a fundamental Higgs, \`a la `finite naturalness'~\cite{Farina:2013mla}, by computing loops involving the heavy fermion as done in~\cite{Davighi:2023iks}).

In the next section we discuss in detail the steps III and IV, which are specific to the 
composite Higgs model. As we shall see, the explicit construction of the Higgs potential, and the requirement of a light Higgs mass, forces us to choose $\langle \Sigma_R \rangle = O(\fHC)  \ll \LambdaHC$. This, in turn, implies that the lightest exotic states of this setup are the massive  $Z^\prime$ and $W^\pm_R$ bosons, plus the top partners, following from the SSB in Eqs.~(\ref{eq:linkVEVs1})--(\ref{eq:linkVEVs2}). Phenomenological  constraints and implications of these fields are discussed 
in \S \ref{sec:pheno}.

\section{Composite dynamics}
\label{sec:composite}

\subsection{Notation and conventions for the Goldstone boson fields}

In our convention, $S p(4) \subset S U(4)$ is the 10-dimensional group of $4 \times 4$ special unitary matrices $\{U\}$ that moreover satisfy $U^T \Omega U=\Omega$, where the symplectic form $\Omega$, in our basis, is given by the following antisymmetric matrix:
$$
\Omega=\begin{pmatrix}
0 & 1 & 0 & 0 \\
-1 & 0 & 0 & 0 \\
0 & 0 & 0 & -1 \\
0 & 0 & 1 & 0
\end{pmatrix}\, .
$$
 
\subsubsection*{$Sp(4)$ Algebra and basis} 

The Lie algebra $\mathfrak{s p}(4)$ and its representations are probably familiar to most readers, thanks to the Lie algebra isomorphism $\mathfrak{s p}(4) \cong \mathfrak{s o}(5)$. The corresponding Lie group isomorphism is $S p(4) \cong \operatorname{Spin}(5)$, where $\operatorname{Spin}(5)$ is the double cover of $S O(5)$ that admits spinor representations.
We choose a particular basis for the fundamental 4-dimensional representation of the $Sp(4)$ group, where the 10 generators take the form 
\be
T^a_L=\frac{1}{2}\left(\begin{array}{cc}
\sigma^a & 0 \\
0 & 0
\end{array}\right)\, , \qquad T^a_R=\frac{1}{2}\left(\begin{array}{cc}
0 & 0 \\
0 & \sigma^a
\end{array}\right)\,, \qquad T_X^{a}=\frac{1}{2 \sqrt{2}}\left(\begin{array}{cc}
0 & \bar{\sigma}^{a} \\
\bar{\sigma}^{a\dagger} & 0
\end{array}\right) \, .
\label{eq:Sp4rep}
\ee
Here $\sigma^a$ denote the Pauli matrices and $\bar{\sigma}^{a}=\left\{ i \sigma^{a}, \mathbb{1}_2 \right\}$, with $\mathbb{1}_2$ being 
the $2\times 2$ identity matrix.\footnote{Lest there is confusion, we use the symbol `$a$' to denote a generic Lie algebra index (usually summed on); {\em e.g.} in the case of $T_L^a$, $a$ runs from $1$ to $3$, while for the broken generators $T_X^a$, $a$ runs from $1$ to $4$.} The $T_{L,R}$ 
correspond to the  $\mathfrak{su}(2)_{L,R}$ sub-algebras of $\mathfrak{sp}(4)$ that we assume to be unbroken, 
while  the $T_X^{a}$ are the broken generators. The generators are normalised such that $\operatorname{Tr}( T^a T^b) = \frac{1}{2} \delta^{ab}$.

\subsubsection*{The Goldstone boson matrix} 
We define the coset element as
\be
U(x) \equiv \exp \left(i\sqrt{2} \phi_a(x) T_X^{a}/\ff \right) \, ,
\label{eq:UUdef}
\ee
where $\phi_{a}(x)$ are the Goldstone bosons and $T_X^{a}$ the broken generators as given above.
The $U(x)$ field transforms under $Sp(4)$ as
\be
U(x)\rightarrow \hat g \cdot U(x) \cdot {\hat h}[\phi_{a};g]^{-1}\,, 
\label{eq:UUtran}
\ee
where $\hat g$ is an element of the $Sp(4)$ group and ${\hat h}[\phi_{a};g]$ is an element of the unbroken subgroup $SU(2)_L\times SU(2)_R^\heavy$. Choosing the representation (\ref{eq:Sp4rep}) for the $Sp(4)$ generators leads to 
\be
U[\phi]=\left(\begin{array}{cc}
\cos (\frac{h}{2 \ff})\, \mathbb{1}_2 & i \sin (\frac{h}{2 \ff})\frac{\PP}{|\PP|} \\
i \sin (\frac{h}{2 \ff}) \frac{\PP ^{\dagger}}{|\PP|} & \cos (\frac{h}{2 \ff})\, \mathbb{1}_2
\end{array}\right)
\equiv 
\left(\begin{array}{cc}
\cos( \frac{\modH}{2 \ff})\, \mathbb{1}_2 & i \sin (\frac{\modH}{2 \ff} )\frac{\PP}{|\PP|} \\
i \sin (\frac{\modH}{2 \ff} )\frac{\PP ^{\dagger}}{|\PP|} & \cos (\frac{\modH}{2 \ff})\, \mathbb{1}_2
\end{array}\right)
\label{eq:Uexp}
\ee
where we have defined the $2\times 2$ matrix
\be \label{eq:Hdefinition}
\PP \equiv \left(\begin{array}{cc}
 i\phi_3 + \phi_4 & i\phi_1 + \phi_2 \\
 i\phi_1 - \phi_2 &  \phi_4 - i\phi_3
\end{array}\right)
\ee 
and the scalar field
\be 
h(x) =  |\PP|
\equiv \sqrt{\mathrm{Det}(\PP)} 
=\sqrt{\phi_1(x)^2 + \phi_2(x)^2 + \phi_3(x)^2+ \phi_4(x)^2}\,.
\ee

Given the factorised structure of the unbroken group, it is convenient to introduce two orthogonal Goldstone bosons {\em vectors}, denoted  $u_{L,R} [\phi]$ and defined as 
\be
u_L[\phi]\equiv U \cdot \mathbb{P}_L =\left(\begin{array}{c}
\cos (\frac{\modH}{2 \ff} )\, \mathbb{1}_2  \\
i \sin( \frac{\modH}{2 \ff} )\frac{\PP^{\dagger}}{|\PP|}  
\end{array}\right)\,, 
\qquad 
u_R[\phi]\equiv U \cdot \mathbb{P}_R=\left(\begin{array}{c} 
i \sin (\frac{\modH}{2 \ff}) \frac{\PP}{|\PP|} \\
 \cos (\frac{\modH}{2 \ff})\, \mathbb{1}_2
\end{array}\right)\,,
\ee
where the 4 $\times$ 2 left and right projectors are  $\mathbb{P}_L \equiv \begin{pmatrix} \mathbb{1}_2 & 0 \end{pmatrix}^T$ and $\mathbb{P}_R \equiv \begin{pmatrix} 0& \mathbb{1}_2  \end{pmatrix}^T$.
Using these vectors the full $U[\phi]$ matrix can be written as $U [\phi]  = ( u_L[\phi] ,  u_R[\phi])$. The transformation 
properties of  $u_{L,R} [\phi]$ under 
$Sp(4)$ assume the following convenient form
\be
 u_L \to  \hat g u_L \hat h_L^\dagger\,, \qquad    u_R \to  \hat g u_R \hat h_R^\dagger\,, \qquad  
\ee
where, for simplicity, we have omitted to indicate the dependence of the $\hat h_{L/R}$ elements
on the Goldstone bosons.
Note also that 
\be
u_L^\dagger  u_R = u_R^\dagger u_L = 0\,, \qquad u_R^\dagger  u_R = u_L^\dagger  u_L = \mathbb{1}_2\,. 
\ee
\subsubsection{Lowest-order Goldstone boson Lagrangian}
\label{sec:GBL}
Using these definitions, the canonically normalised Lagrangian at leading order in the derivative expansion,
including external gauge fields of $SU(2)_L \times SU(2)^\heavy_R$,
can be written as
\be
\label{kinU}
\mathcal{L}_U^{(2)}=\frac{\ff^2}{2} \operatorname{Tr}_{[4]} \left[ ( \hat D ^\mu U)^\dagger \hat D_\mu U \right]\,,
\ee
where 
\be
\hat D_\mu U = \partial_\mu U -  i U \hat \Gamma_\mu  -ig_L \left[\hat{A}^L_\mu,U\right]-ig_R\left[\hat{A}^R_\mu,U\right]~.
\ee
Following the formalism of Callan, Coleman, Wess, and Zumino (CCWZ)~\cite{Callan:1969sn}, we define $\hat \Gamma_\mu$ from the decomposition of $U^\dagger \partial_\mu U$ in terms 
of broken and unbroken generators,
\be
\label{uderu}
U^\dagger \partial_\mu U = i (\Gamma_\mu^a T_L^a+\Gamma_\mu^a T_R^a) + i u^a_\mu T^a_X \equiv  i \hat \Gamma_\mu + i \hat u_\mu, 
\ee
such that under {\em global} $Sp(4)$ transformations
\bea
\hat \Gamma_\mu &\to&   \hat h[\phi_a; g] \Gamma_\mu  \hat h[\phi_a; g]^\dagger -  i \hat h[\phi_a; g]\, \partial_\mu \hat h^\dagger [\phi_a; g]\,, \no\\
\hat u_\mu &\to&   \hat h[\phi_a; g] \hat u_\mu  \hat h[\phi_a; g]^\dagger \,.
\eea
On the other hand, $\hat{A}^{L,R}_\mu$ and $g_{L(R)}$ denote gauge fields and related couplings ensuring {\em local} invariance under $SU(2)_L \times SU(2)^\heavy_R$.
Note that $\hat\Gamma_\mu$ is a function of the Goldstone bosons, but its expansion in powers of $\phi_a$ starts at second order, hence it plays no role in the expansion of $\mathcal{L}_U^{(2)}$ up to quadratic terms in $H$ (see appendix~\ref{sec:Gamma} for more details). 
As we shall see, the constant $\ff$ appearing in the normalisation of  $\mathcal{L}_U^{(2)}$ and in the expansion of $U[\phi]$ in 
(\ref{eq:Uexp}) is related to the 
order parameter of the strong sector $\fHC$,
defined as in~\cite{Giudice:2007fh}, via 
$\ff=\fHC/2$.

From \eqref{kinU}, expanding up to second order in  $\PP$, which is equivalent to considering the $F\to\infty$ limit,
we can derive the dimension-4 part of the effective  Lagrangian describing the dynamics of the field  $\PP$, which transforms as a $\left(\mathbf{2},\mathbf{2}\right)$ under $SU(2)_L \times SU(2)^\heavy_R$:
\be
\label{ourH}
\cL^{(2)}_H = \frac{1}{4} \operatorname{Tr}_{[2]}\left(D^\mu \PP \right)^\dagger \left( D_\mu \PP\right)\,.
\ee
In this case the covariant derivative reads 
\be \label{eq:cov-deriv-Phi}
D_\mu  \PP = \partial_\mu \PP - i g_L  \hat A^L_\mu \PP + i g_R  \PP  \hat A^R_\mu \, 
\qquad \hat A^{L(R)}_\mu = \frac{1}{2}  \sigma^a  A^a_{L(R), \mu}~.
\ee
At this point it is clear that
in the limit $F\to\infty$ we can identify $\PP$ with the SM Higgs field
(in a bi-doublet notation), and $h(x)$ with its radial component:
\be
\PP (x)|_{\rm unit.-gauge} =  h(x)\, \mathbb{1}_2\,,
\ee
such that the vacuum expectation value of $\PP$ reads 
\be
\label{VEVh}
\langle \PP \rangle = \langle  h \rangle\,  \mathbb{1}_2  \equiv  v\, \mathbb{1}_2 \, .
\ee
Considering only 
$\cL^{(2)}_H$ in (\ref{ourH}),  the value of 
$v$ thus defined yields $m^2_W =g_L^2 v^2/4$.

\subsubsection*{Casimirs of  $SU(2)_L \times SU(2)^\heavy_R$ and $SU(2)^\heavy_R$ to encode explicit
breaking}

It is convenient to introduce the following quadratic
Casimir elements of the unbroken sub-algebra $\mathfrak{su}(2)_L \oplus \mathfrak{su}(2)_R^\heavy$, written in the fundamental representation of $\mathfrak{sp}(4)$:
$$
\Delta_R \equiv \Delta=\left(\begin{array}{cc}
0 & 0 \\
0 & \mathbb{1}_2
\end{array}\right), \qquad \Delta_L \equiv \mathbb{1}_4-\Delta=\left(\begin{array}{cc}
\mathbb{1}_2 & 0 \\
0 & 0
\end{array}\right)\,.
$$
Any linear combination of these two matrices (excluding $\Delta_R+\Delta_L=\mathbb{1}_4$) commutes with the unbroken generators, and does not commute with the broken ones:
\be
\left[\Delta_{L,R}, T^a_{L,R}\right]=0 \qquad\left[\Delta_{L,R}, T_X^{a}\right] \neq 0\, .
\ee
In building effective operators, we can treat 
$\Delta_{L,R}$ as
{\em spurions} of $Sp(4)$, in the sense that these matrices can be used to encode the effects of explicit $Sp(4)$ breaking present in our model.

We remark that $\Delta_{L,R}$ do {\em not}  correspond strictly to the VEVs of some (even spurious) fundamental field transforming in a linear representation of $\mathcal{G}$, but can (in certain cases, such as when $\Delta$ appear below to encode the effects of gauging) arise from integrating out more complicated combinations of UV fields. In any case, identifying $\Delta_{L/R}$ as quadratic Casimir elements in the universal enveloping algebra, we can take them 
formally to transform under $\mathcal{G}$ as
\be
\Delta_{(L, R)} \rightarrow \hat g \Delta_{(L, R)} \hat g^{-1}\,.
\label{eq:spurionsT}
\ee
This allows us to systematically build 
operators that break $Sp(4)$ but are invariant under 
 $SU(2)_L \times SU(2)^\heavy_R$, keeping track of the origin of this explicit   symmetry breaking. Since $\Delta_L+\Delta_R=\mathbb{1}_4$, 
the two spurions are not independent and we can limit ourselves to introduce a single one that we choose to be 
$\Delta \equiv \Delta_R$.

Since we are interested in describing also the explicit breaking $SU(2)_R^\heavy \to U(1)^\heavy_{T^3_R}$, in particular to explain the splitting between the top and bottom quark masses, we further introduce the spurion
\be
\Delta_\Sigma = \left(\begin{array}{cc}
0 & 0 \\
0 &  \sigma_3 \end{array}\right)\,, \qquad 
\label{eq:spurions}
\ee
with formal $Sp(4)$ 
transformation properties  as in (\ref{eq:spurionsT}).
By construction, $\Delta_\Sigma$
commutes only with the generators of
 $SU(2)_L \times U(1)^\heavy_R$. Note that, since $\Delta_\Sigma$ `explicitly breaks' $SU(2)_R^\heavy$, which recall is gauged, this really means that operators built in terms 
of $\Delta_\Sigma$ in the composite sector effectively describe 
the result of the non-vanishing expectation value of $\Sigma_R$ 
in (\ref{eq:linkVEVs1}), after integrating  out heavy fields that couple to $\Sigma_R$. 

\subsection{Coupling to fermions}

\subsubsection{Third-family partial compositeness} 
\label{sect:topY}
As anticipated, we assume that the coupling of the (elementary) third-generation fermions to the composite Higgs  is the  result of a linear mixing between the elementary fermions and appropriate fermionic operators of the strong sector. For simplicity, we illustrate this mechanism in the quark sector, denoting generically with $\mathcal{O}_q$ the 
corresponding fermionic operators.
To build invariant terms, it  is more convenient to use  
$u_L$ and $u_R$ rather than the $U$ field. The most general linear mixing is described by 
\be
\label{partcompfull}
\mathcal{L} \supset \ff \left[\bar{q}^\heavy_L u_L^{\dagger}\left(\lambda^q_L \mathbb{1} +\tilde{\lambda}^q_L \Delta\right) \mathcal{O}_q +\bar{q}^\heavy_R u_R^{\dagger}\left(\lambda^q_R \mathbb{1} +\tilde{\lambda}^q_R \Delta\right) \mathcal{O}_q \right] +{\rm h.c.},
\ee
where $\lambda^q_{L(R)}$ and $\tilde{\lambda}^q_{L(R)}$ are arbitrary complex couplings.
The terms proportional to $\lambda^q_{L(R)}$ preserve $Sp(4)$-invariance, while those
proportional to $\tilde{\lambda}^q_{L(R)}$ do not: the spurion matrices $\Delta$ appear in this context to act as projectors onto relevant fermionic degrees of freedom. The explicit breaking introduced by these latter terms is essential to generate a SM-like effective Yukawa interaction.
To ease the notation, in the following we omit the flavour and gauge
indices on the elementary fields. 

\begin{figure}[t]
\begin{center}
\begin{tikzpicture}
\begin{feynman}
\vertex (a) { $\overline{q}^\heavy_{L}$};
\vertex [right=0.8in of a] (b);
\vertex [right=0.6in of b] (c);
\vertex [right=0.6in of c] (d);
\vertex [right=0.6in of d] (e) { $q_{R}^\heavy$};
\node at (b) [circle,fill,inner sep=1.5pt,label=below:{ $\{\lambda, \tilde\lambda\}_{L}\,\,$}]{};
\node at (d) [circle,fill,inner sep=1.5pt,label=below:{ $\{\lambda, \tilde\lambda\}_{R}\,\,$}]{};
\node at (c) [square dot,fill,inner sep=1.0pt]{};
\vertex [above=0.7in of b] (f) { $u_L^\dagger$};
\vertex [above=0.7in of d] (g) { $u_R$};
\diagram* {
(a) -- [plain] (b)  -- [double,edge label={ $\mathcal{O}_q$}] (c) -- [double,edge label={ $\overline{\mathcal{O}}_q$}] (d) -- [plain] (e),
(b) -- [scalar] (f),
(d) -- [scalar] (g),
};
\end{feynman}
\end{tikzpicture}
\end{center}
\caption{Diagram describing the mechanism responsible for  third-generation Yukawa couplings.
 \label{fig:VLF} }    
\end{figure}
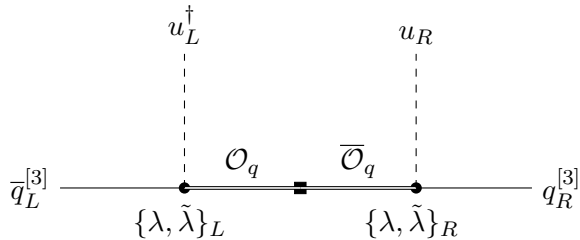

Integrating out the heavy composite states as 
shown schematically in Fig.~\ref{fig:VLF}, and 
assuming that the  spectrum of the composite 
states is characterized by a single mass scale $M_q$,
the effective Yukawa interaction among elementary 
fields assumes the form 
\be
\mathcal{L}_Y^{\rm eff}  =  \lambda^q_L \lambda^{q*}_R \kappa^q_{LR}  
\frac{F^2}{M_q}  \, \overline{q}_L  u_L^\dagger \Delta u_R^{\phantom{\dagger}} q_R 
 = \frac{y_q}{\sqrt{2}} \, \overline{q}_L H q_R + O(H^2)\,,
 \label{eq:Yeff}
\ee
where 
\be
  |y_q| = \lambda^q_L \lambda^{q*}_R \kappa^q_{LR} 
\frac{F} {\sqrt{2} M_q}\,, 
\qquad 
 \kappa^q_{LR} =\left(1+\tilde{\lambda}^q_L / \lambda^q_L\right)\left(1+\tilde{\lambda}^{q*}_R / \lambda^{q*}_R\right)-1\,.
\ee
As can be seen, $y_q$ is non-vanishing if both   $q_{L}$ and $q_{R}$
mix with the composite states, and at least one of the two mixing terms is not 
$Sp(4)$ invariant.\footnote{In principle, we could also introduce  $Sp(4)$-breaking terms in the mass-matrix of the composite states. %The composite states  belongs to complete $Sp(4)$ representations and, in absence of explicit breaking, exhibit mass degeneracy in each representation.
However, this is a subleading effect  with respect to the  breaking present in the linear couplings of the elementary fields to the composite sector, hence we neglect it in the following.}

The effective Yukawa interaction in 
Eq.~(\ref{eq:Yeff}) is $SU(2)^\heavy_R$ invariant. The breaking of custodial symmetry in the Yukawa sector, which is necessary to describe the top-bottom mass splitting, can be achieved adding terms proportional to $\Delta_\Sigma$  in  
Eq.~(\ref{partcompfull}). 
The limiting case where only the top quark Yukawa coupling is non-zero can be obtained, in particular,   via the replacement 
\be
 \Delta \to \Delta_+ = \frac{1}{2} (\Delta+\Delta_\Sigma)
\equiv
  \left(\begin{array}{cc}
0 & 0 \\
0 &  \sigma_+ \end{array}\right)\,.  
\ee 
We emphasize that, since the $\Delta_\Sigma$ part breaks $SU(2)_R^\heavy$, it must in reality be built from insertions of the $SU(2)_R^\heavy$-breaking VEV of $\Sigma_R$ (with the $SU(2)_R^\light$ indices contracted appropriately). In this sense, the near cancellation of the bottom Yukawa (relative to the top) is a result of fine-tuning between two independent quantities.\footnote{One can avoid this tuning by a variation of the model in which only a $U(1)_R^\heavy \subset SU(2)_R^\heavy$ subgroup of the global symmetry is gauged. Then $t_R$ and $b_R$ are two independent fields, from which one can construct two independent sets of (separately) gauge-invariant operators that give rise to independent top and bottom Yukawa couplings after the strong sector symmetry breaking transition. }

\subsubsection{Fermion contribution to the Higgs potential} \label{sec:fermion}

In the absence of explicit $Sp(4)$ breaking, the  Higgs field would be an exact Goldstone boson and therefore massless.  The explicit breaking of $Sp(4)$ induced by the couplings of the elementary fermions to the composite sector, together with the gauging of only a subgroup of $Sp(4)$ (soon to be discussed), transform the Higgs field into a pseudo Goldstone boson
with non-vanishing Higgs potential.
In this section we evaluate the one-loop  contribution to the potential obtained by integrating out the elementary fermions. 

Following the Coleman--Weinberg approach, the starting point is to identify the effective vertices with two fermion fields and  arbitrary powers of the Higgs field at zero 
momentum transfer. We parameterise these two-point functions via appropriate form factors which depends only on the momentum of the fermion fields. Taking into account only the spurion terms necessary to generate the top-quark Yukawa coupling, we define 
\bea
\mathcal{L}_{\rm eff} &\supset & \overline{q}_L \slashed{p} \left[\Pi^{q_L}_0(p^2) \mathbb{1} + \Pi^{t_L}_1(p^2) u_L^\dagger \Delta_+ u_L\right] q_L +
\overline{q}_R \slashed{p} \left[\Pi^{q_R}_0(p^2) \mathbb{1} + \Pi^{t_R}_1(p^2) u_R^\dagger \Delta_+ u_R\right] q_R 
\nonumber \\ 
&&+ \left\{ \overline{q}_L \left[ \mathcal{M}_t(p^2) u_L^\dagger \Delta_+ u_R\right] q_R
+ {\rm h.c.} \right\}\,.
\label{eq:M1}
\eea
Omitting to indicate the momentum dependence of the form factors, and evaluating the 
explicit dependence from the Higgs field in the unitary gauge, leads to 
\begin{align}
\mathcal{L}_{\text {eff }} \supset \, \, &\overline{t}_L \slashed{p}\left[ \Pi_0^{q_L}+\Pi_1^{t_L} \sin^2\left(\frac{\modH}{2F}\right)\right] t_L  +\overline{t}_R \slashed{p}\left[\Pi_0^{t_R}-\Pi_1^{t_R} \sin^2 \left(\frac{\modH}{2F}\right)\right] t_R\nonumber\\ + &\left\{  \overline{t}_L\left[ \mathcal{M}_t \sin \left(\frac{\modH}{2F}\right) \cos \left(\frac{\modH}{2F}\right)   \right]t_R   
+ {\rm h.c.} \right\}\,,
\end{align}
where $\Pi_0^{t_R}=\Pi_0^{q_R}+\Pi_1^{t_R}$. Note that we moved from a doublet notation for the fermions in (\ref{eq:M1}) to an explicit indication of the individual components
($t_L$ and $t_R$) coupled to the Higgs field in the unitary gauge. 

The value of the form factors at $p^2=0$ can be either fixed by the normalization of the fields or expressed in terms of parameters controlling the elementary-composite fermion mixing.  In particular, via a field redefinition we can set 
\be
\Pi_0^{q_L}(0)=
\Pi_0^{t_R}(0)=1\,,
\ee
while the matching with (\ref{partcompfull})
leads to 
\be
\label{expformpsmall}
\Pi^{t_{L}}_1(0) = \frac{F^2}{M_T^2} 
\left[2\lambda^t_{L}\tilde{\lambda}^t_{L}+ (\tilde{\lambda}^t_{L})^2 \right] \equiv \frac{F^2}{M_T^2} (\lambda^t_{L})^2 \kappa^t_{L}
\,, \qquad 
\Pi^{t_{R}}_1(0) \equiv \frac{F^2}{M_T^2} (\lambda^t_{R})^2 \kappa^t_{R}
\,,
\ee
and 
\be
|\mathcal{M}_t(0) |  =  \frac{F^2}{M_T} \lambda^t_L \lambda^t_R \kappa^t_{LR} \equiv y_t \sqrt{2} F\,.
\ee
Here we have denoted with $M_T$ the effective mass of the composite state(s) mixing with the top quark (the so-called `top partners'), while $y_t$ denotes the SM-like Yukawa coupling 
($y_t = \sqrt{2}m_t/v \approx 1$).

\begin{figure}[t]
\centering
 \begin{subfigure}[t]{0.31\textwidth}
\centering
\begin{tikzpicture}[baseline=(b.base)]
  \begin{feynman}
    \vertex (a);
    \vertex [above left=of a] (i1);
    \vertex [below left=of a] (i2);
    \vertex [left=0.35cm of a] (k1){\small $\Pi_1^{t_{L(R)}}$};
    \vertex [right=0.7cm of a] (c);
    \vertex [above= 0.6 cm of c] (l1){\small $t_{L(R)}$};
    \vertex [right=1.25cm of c] (d){+};
    \vertex [right=1.4cm of a] (b);
    \diagram* {
(a) -- [half left] (b),
(b) -- [half left] (a),
      (i1) -- [dashed] (a) -- [dashed] (i2),
    };
    \node [draw=black, circle, fill=black, inner sep=0.7mm] at (a) {};
  \end{feynman}
\end{tikzpicture}
 \end{subfigure}
 \begin{subfigure}[t]{0.2\textwidth}
\centering
\begin{tikzpicture}[baseline=(b.base)]
  \begin{feynman}
    \vertex (a);
    \vertex [above left=of a] (i1);
    \vertex [below left=of a] (i2);
    \vertex [above right=of b] (j1);
    \vertex [left=0.35cm of a] (k1){\small $\mathcal{M}_t$};
    \vertex [above= 0.6 cm of c] (l1){\small$t_L$};
    \vertex [below= 0.6 cm of c] (l2){\small$t_R$};
    \vertex [right=0.35cm of b] (k2){\small $\mathcal{M}_t$};
    \vertex [below right=of b] (j2);
    \vertex [right=0.7cm of a] (c);
    \vertex [right=1.88cm of c] (d){+ \,\,\,\ldots};
    \vertex [right=1.4cm of a] (b);
    
    \diagram* {
(a) -- [half left] (b),
(b) -- [half left] (a),
      (i1) -- [dashed] (a) -- [dashed] (i2),
      (j1) -- [dashed] (b) -- [dashed] (j2),
    };
    \node [draw=black, circle, fill=black, inner sep=0.7mm] at (a) {};
\node [draw=black, circle, fill=black, inner sep=0.7mm] at (b) {};
  \end{feynman}
\end{tikzpicture}
 \end{subfigure}
 
 \bigskip
 
\begin{subfigure}[b]{0.23\textwidth}
\centering
\begin{tikzpicture}[baseline=(b.base)]
  \begin{feynman}
    \vertex (a);
    \vertex [above left=of a] (i1);
    \vertex [below left=of a] (i2);
    \vertex [above= 0.6 cm of c] (l1){\small $A_{L(R)}$};
    \vertex [left=0.25cm of a] (k){\small $\Pi_1$};
    \vertex [right=0.7cm of a] (c);
     \vertex [right=1.25cm of c] (d){+};
    \vertex [right=1.4cm of a] (b);
    \diagram* {
(a) -- [boson, half left] (b),
(b) -- [boson, half left] (a),
      (i1) -- [dashed] (a) -- [dashed] (i2),
    };
    \node [draw=black, circle, fill=black, inner sep=0.7mm] at (a) {};
  \end{feynman}
\end{tikzpicture}
 \end{subfigure}
 \begin{subfigure}[b]{0.32\textwidth}
\centering
\begin{tikzpicture}[baseline=(b.base)]
  \begin{feynman}
    \vertex (a);
    \vertex [above left=of a] (i1);
    \vertex [below left=of a] (i2);
    \vertex [above right=of b] (j1);
    \vertex [above= 0.6 cm of c] (l1){\small $A_{L}$};
    \vertex [below= 0.6 cm of c] (l2){\small $A_{R}$};
    \vertex [left=0.25cm of a] (k1){\small $\Pi_1$};;
    \vertex [right=0.25cm of b] (k2){\small $\Pi_1$};;
    \vertex [below right=of b] (j2);
    \vertex [right=0.7cm of a] (c);
     \vertex [right=2cm of c] (d){+};
    \vertex [right=1.4cm of a] (b);
    
    \diagram* {
(a) -- [boson, half left] (b),
(b) -- [boson, half left] (a),
      (i1) -- [dashed] (a) -- [dashed] (i2),
      (j1) -- [dashed] (b) -- [dashed] (j2),
    };
    \node [draw=black, circle, fill=black, inner sep=0.7mm] at (a) {};
\node [draw=black, circle, fill=black, inner sep=0.7mm] at (b) {};
  \end{feynman}
\end{tikzpicture}
 \end{subfigure}
  \begin{subfigure}[b]{0.2\textwidth}
\centering
\begin{tikzpicture}[baseline=(b.base)]
  \begin{feynman}
    \vertex (a);
    \vertex [above left=of a] (i1);
    \vertex [below left=of a] (i2);
    \vertex [left=0.25cm of a] (k1){\small$\Pi_1$};
    \vertex [right=0.25cm of b] (k2){\small$\Pi_\Sigma$};
    \vertex [above right=of b] (j1) {\small$\Sigma^\dagger$};
    \vertex [below right=of b] (j2){\small$\Sigma$};
    \vertex [right=0.7cm of a] (c);
    \vertex [above= 0.6 cm of c] (l1){\small $A_R$};
    \vertex [below= 0.6 cm of c] (l2){\small$A_R$};
     \vertex [right=1.8cm of c] (d){+ \,\,\,\ldots};
    \vertex [right=1.4cm of a] (b);
    
    \diagram* {
(a) -- [boson, half left] (b),
(b) -- [boson, half left] (a),
      (i1) -- [dashed] (a) -- [dashed] (i2),
      (j1) -- [dashed] (b) -- [dashed] (j2),
    };
    \node [draw=black, circle, fill=black, inner sep=0.7mm] at (a) {};
\node [draw=black, thick, circle, fill=black, inner sep=0.7mm] at (b) {};
  \end{feynman}
\end{tikzpicture}
 \end{subfigure}
\caption{Schematic illustration of the one-loop contributions to the Higgs potential, coming from the explicit $Sp(4)$ breaking required in both the fermion (\ref{sec:fermion}) and gauge (\ref{sec:gauge}) sectors. 
 \label{Fig:pot1}}
\end{figure}
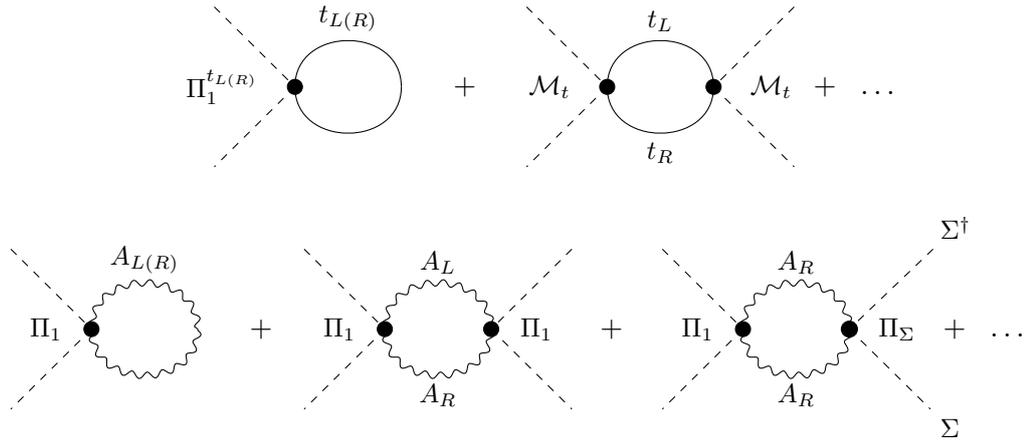

To compute the one-loop  potential, we need to resum the series of one-loop diagrams involving arbitrary number of insertions of form factors, as indicated schematically in Fig.~\ref{Fig:pot1}.  The result thus obtained is 
\begin{align}
\Delta V(h)_{f} = -2N_c & \int \frac{d^4 p_E}{(2 \pi)^4}\left\{\log \left[1 +\frac{\Pi_1^{t_L}}{\Pi_0^{q_L}} \sin^2\left(\frac{\modH}{2F}\right)\right] + \log\left[ 1-\frac{\Pi_1^{t_R}}{\Pi_0^{t_R}} \sin^2\left(\frac{\modH}{2F}\right)\right]\right.\nonumber \\
+ & \left.\log \left[ 1+\frac{ |\mathcal{M}_t|^2 \sin^2 \left(\frac{\modH}{2F}\right) 
\cos^2 \left(\frac{\modH}{2F}\right) }{p_E^2\left(\Pi_0^{q_L}+\Pi_1^{q_L}\sin^2\left(\frac{\modH}{2F}\right) \right)\left(\Pi_0^{q_R}-\Pi_1^{q_R}\sin^2\left(\frac{\modH}{2F}\right) \right)}\right]\right\}, 
\label{eq:Vf}
\end{align}
where the pre-factor is determined by the number of colors
and the two spin degrees of freedom of the fermion fields. The explicit evaluation of the integral, which requires one to choose a functional dependence for the form factors at $p^2 \not=0$, is discussed in \S \ref{sect:HiggsPot}.

\subsection{Coupling to gauge bosons}

\subsubsection{Chiral structure}

In this subsection, the flavour deconstruction of the electroweak gauge group will play a crucial role when we calculate the impact of gauging on the pNGB Higgs potential.

As for the fermion sector, in order to compute the contribution of the $SU(2)_L \times SU(2)_R^\heavy$ gauge fields to the  Higgs potential we need to determine the two-point function of two 
gauge fields and an arbitrary number of Higgs fields. The mass term can be derived directly from the chiral Lagrangian, where the symmetry breaking pattern $Sp(4) \to SU(2)_L \times SU(2)_R^\heavy$ implies that only one specific combination of $A^L_\mu$ and $A^R_\mu$ appears. This can be easily seen by expanding the covariant derivative acting on $U$ from \eqref{kinU}. In the unitary gauge, the leading and universal term quadratic in $A^L_\mu$ and $A^R_\mu$ is:
\be
\label{covdevgauge}
\left. \cL^{(2)}_U \right|_{{\rm unit.-gauge},~A^2}  = \frac{F^2}{2}  \left( g_L  A^a_{L,\mu} - g_R A^a_{R,\mu}\right)^2  \sin^2 \left(\frac{\modH}{2 F}\right)~.
\ee
This term accounts for the SM relation between the $W$-boson mass and the Higgs VEV by considering the limit $h=v$. 

In principle, additional terms can be constructed by considering non-minimal operators in the chiral Lagrangian, obtained with appropriate insertions of the $\Delta$ matrices. 
In this context (namely, of encoding the effects of gauging a non-trivial subgroup), the $\Delta$ matrices can be thought of as arising from tracing over a quadratic form built from a fundamental spurion field that transforms in the adjoint representation of $\mathcal{G}=Sp(4)$, with non-vanishing components only in directions that pick out the unbroken $\mathcal{H}$ generators.
For example, the non-minimal term 
\be
 \cL^{(2')}_U  =  \delta_\pi F^2 \,  \operatorname{Tr}_{[4]}\left[ \Delta ( \hat D_\mu U^\dagger U) \Delta  ( U^\dagger \hat D^\mu U) \right]
\ee
 leads to 
\be
\left. \cL^{(2')}_U \right|_{{\rm unit.-gauge},~A^2}  = \delta_\pi   \frac{ F^2 }{ 2 } \left( g_L  A^a_{L,\mu} - g_R A^a_{R,\mu}\right)^2  \sin^4 \left(\frac{\modH}{2 F}\right) \, .
\label{covdevgaugeNLO}
\ee
Again, the same relative sign between $SU(2)_L$ and $SU(2)_R^\heavy$ gauge bosons appears, but now with a different trigonometric dependence on $h/2F$. The required insertion of the spurions in $\cL^{(2')}_U$ 
implies this interaction term is suppressed compared to $\cL^{(2)}_U$ ({\em i.e.}~we expect $|\delta_\pi|\ll1$). Hence higher powers of trigonometric functions in the two-point function of the gauge fields are expected to be suppressed, even though both terms (\ref{covdevgauge}) and (\ref{covdevgaugeNLO}) are formally the same order ({\em i.e.} quadratic) in the gauge couplings $g_{L/R}$ which here act as the spurion couplings.\footnote{The fact that there are two independent functions $\sin^2(h/2F)$ and $\sin^4(h/2F)$ that appear in the Lagrangian at leading (quadratic) order in the gauge couplings corresponds to the group-theoretic fact that there are two independent quadratic Casimir elements in the unbroken subgroup. This counting, as described in Ref.~\cite{Gripaios:2015qya}, corresponds to the pair of real irreps ${\bf (3,1)}$ and ${\bf (1,3)}$ of $\mathcal{H}=SU(2)_L \times SU(2)_R^\heavy$ that appear in the decomposition of the adjoint of $Sp(4)$ to $\mathcal{H}$ (after discarding the adjoint of $\mathcal{H}$ itself).
}

\subsubsection{Gauge boson contribution to the Higgs potential}
\label{sec:gauge}

Taking into account also the kinetic terms of $A^L_\mu$ and $A^R_\mu$, the complete 
two-point function written in terms of form factors for arbitrary momenta of the gauge fields
  assumes the form
\bea
&& \mathcal{L}^{(A^2)}_{\rm eff} =\frac{1}{2}  (P_T)^{\mu \nu} \Big\{  \Pi_0 \left(q^2\right) \left[ A^a_{L,\mu} A^a_{L,\nu} + A^a_{R,\mu} A^a_{R,\nu}  \right]  
\label{eq:LAA0}\\
&& + \left[ \Pi_1  \left(q^2\right)\sin^2 \left(\frac{\modH}{2F}\right)+\Pi_2 \left(q^2\right)\sin^4 \left(\frac{\modH}{2F}\right) \right]  \left( g_L  A^a_{L,\mu} - g_R A^a_{R,\mu} \right) \left( g_L  A^a_{L,\nu} - g_R A^a_{R,\nu} \right) \Big\}  \nonumber \,,
\eea
where the transverse projector is $(P_T)^{\mu \nu} \equiv \eta_{\mu \nu} - q_\mu q_\nu /q^2$. Matching to the usual Yang-Mills Lagrangian, we identify $\Pi_0 (q^2) = -q^2$, up to (irrelevant) higher order terms in powers $q^2$. From \eqref{covdevgauge} and
\eqref{covdevgaugeNLO},
we deduce the normalization conditions for $\Pi_1$ and  $\Pi_2$ in the low momentum limit:
\be
\label{normpi1}
\Pi_1(0) = F^2\,, \qquad
\Pi_2(0) = \delta_\pi F^2\,.
\ee

In the case of the right-handed fields an additional contribution to the two-point function 
is generated by the kinetic Lagrangian of  $\Sigma_R$: when this link field acquires a VEV,
a mass term for the $A^a_{R,\mu}$ field is generated.  In the limit where we neglect the $U(1)_Y^\light$ gauge coupling
(see \S \ref{sec:flavoredGB}), this effect is described by 
 \be
 \Delta \mathcal{L}^{(A_R^2)}_{\rm eff} =\frac{1}{4}  (P_T)^{\mu \nu}   g_R^2
  v^2_\Sigma  \,  A^a_{R,\mu}  A^a_{R,\nu} \,. 
\label{eq:AffR}
\ee
This term does contain any 
coupling to the Higgs field; however, it affects the calculation of the Higgs potential since it modifies the propagator of the right-handed gauge fields. 
In analogy with \eqref{eq:LAA0}, it turns out to be convenient to define $\Pi_\Sigma = v_\Sigma^2/2$.

Similarly to the fermion sector, to compute the Coleman--Weinberg potential we need to sum all 
the one-loop bubbles of gauge fields with arbitrary insertions of the form factors (see Fig.~\ref{Fig:pot1}).
The contribution to the potential thus obtained is:
\begin{align}
\label{eq:VA}
\Delta V(h)_A = +\frac{9}{2} \int dq_E^4 \,\,\, &\log \left(1+g_L^2\frac{\Pi_1 \left(q_E^2\right)\sin^2  \left(\frac{\modH}{2F}\right)+\Pi_2 \left(q_E^2\right)\sin^4  \left(\frac{\modH}{2F}\right)}{\Pi_0^L\left(q_E^2\right)}\right)\\+&\log \left(1+g_R^2\frac{\Pi_1 \left(q_E^2\right)\sin^2  \left(\frac{\modH}{2F}\right)+\Pi_2 \left(q_E^2\right)\sin^4  \left(\frac{\modH}{2F}\right)+ \Pi_\Sigma }{\Pi_0^R\left(q_E^2\right)}\right)\nonumber \\+&\log \left(1+g_Lg_R\frac{4\left(\Pi_1\left(q_E^2\right)\sin^2  \left(\frac{\modH}{2F}\right)+\Pi_2\left(q_E^2\right)\sin^4  \left(\frac{\modH}{2F}\right)\right)^2}{\tilde{\Pi}_0^2\left(h,q_E^2\right) }\right)\nonumber\,,
\end{align}
where
\be 
\tilde{\Pi}_0\left(\modH,q^2\right) \equiv \Pi_0 \left(q^2\right) +\Pi_1 \left(q^2\right)\sin^2  \left(\frac{\modH}{2F}\right)+\Pi_2 \left(q^2\right)\sin^4  \left(\frac{\modH}{2F}\right)\, .
\ee
The numerical pre-factor of $\frac{9}{2}$ originates from the fact that there are three Lorentz polarizations for the vector field, that there are three $SU(2)_{L,R}$ degrees of freedom, and there is a relative factor of $\frac{1}{2}$ with respect to the fermion contribution due to the fact that the gauge bosons are real and not complex. 

\subsection{The Higgs potential}
\label{sect:HiggsPot}

We can now compute explicitly the radiatively induced Higgs potential, 
combining both the fermion and gauge contributions that we have just discussed.

In both cases, the UV behaviour of the relevant form factors should be modelled to ensure a convergence of the corresponding integrals, such that the 
the logarithms in \eqref{eq:Vf} and \eqref{eq:VA} 
are well approximated by their expansion to the first or  
second order (according to the parametric dependence on the couplings).  This implies a general decomposition of the type
\be
V(h) = \Delta V_f(h) +\Delta V_A(h) \approx 
 c_0 -  c_1 \sin^2 \left(\frac{\modH}{2F}\right) + c_2 \sin^4 \left(\frac{\modH}{2F}\right)\,.
\label{eq:Vh}
\ee
Higher-power trigonometric functions,  such as $\sin^6 (\modH/2F)$, can be safely neglected.\footnote{We remark that it has recently been observed that generating the Higgs potential at higher-order, which is radiatively stable if the symmetry breaking spurion is in a large representation, can offer an alternative route to avoiding the $v/F$ tuning~\cite{Durieux:2021riy,Durieux:2022sgm}. }
Before proceeding with the calculation of the fermion and gauge contributions to  $c_{1,2}$, note that na\"ive dimensional analysis implies $c_{1,2} = O(1) \times F^4$, while the physical conditions
to be imposed in order to recover the SM potential at leading order are 
\be
\left. \frac{c_1}{F^4} \right|_{\rm phys.}  =  \frac{m_h^2}{F^2} \qquad \text{ and } \qquad 
\left. \frac{c_2}{F^4} \right|_{\rm phys.}  =  \frac{ 2m_h^2}{v^2} \approx \frac{1}{2}\,.
\label{eq:c12phys}
\ee
 Hence $c_2$ has a natural value. On the other hand, the experimental bounds on $F$  (see \S \ref{sec:Fbounds} and \S \ref{sec:topp_pheno}) require  $m^2_h/F^2 \lesssim 0.03$, which necessarily implies a sizable tuning in $c_1$.

To obtain explicit expressions for 
$c_{1,2}$ in terms of the model parameters, we assume the following simple functional form for the fermion form factors
\be
\mathcal{M}_t(q^2) = \mathcal{M}_t(0) \times \frac{ M_T^2}{ M_T^2 -q^2 } 
\ee
and 
\be
 \frac{ \Pi^{t_{L}}_1(q^2) }{ \Pi^{t_{L}}_1(0) }
 \frac{  \Pi^{q_{L}}_0(0) }{   \Pi^{q_{L}}_0(q^2) }
 =
  \frac{ \Pi^{t_{R}}_1(q^2) }{ \Pi^{t_{R}}_1(0) }
 \frac{  \Pi^{q_{R}}_0(0) }{   \Pi^{q_{R}}_0(q^2) }
 =
  \frac{ M_T^2}{ M_T^2 -q^2 } \frac{ M_f^2}{ M_f^2 -q^2 }\,.
\ee 
The dependence on $M_T$ follows from the assumption of a single top-partner (of mass $M_T$) dominating the diagram in Fig.~\ref{fig:VLF} at small $q^2$. On the other hand, we let $M_f \lesssim 4\pi \fHC$ denote the mass of generic heavy spin-1/2 resonances, which act to cut-off the quadratically divergent loop integrals in \eqref{eq:Vf}.
Similarly, in the gauge sector we assume 
\be
\Pi_{1(2)}(q^2) = \Pi_{1(2)} (0) \frac{M_\rho^2}{M_\rho^2-q^2} \,,
\ee
where $M_\rho \lesssim  4\pi \fHC$ denotes the mass of the 
spin-1 resonances which tame the quadratically divergent integrals in \eqref{eq:VA}. Residual logarithmic divergences in the integrals proportional to $M_f^2$ and $M_\rho^2$ are reabsorbed via 
an $O(1)$ redefinition of these masses.

Under these assumptions,
we find the following expressions for the 
coefficient of the $\sin^2(h/2F)$ term,
\be
\frac{c_1}{F^4} = \left(\frac{c_1}{F^4}\right)^{f-{\rm quad}}
+ \left(\frac{c_1}{F^4}\right)^{\rm min}\,,
\ee
\bea
\left(\frac{c_1}{F^4}\right)^{f-{\rm quad}} &=&
\frac{N_c}{8\pi^2}\left[(\lambda^t_R)^2\kappa^t_{R}-(\lambda^t_L)^2\kappa^t_{L}\right] \frac{ M^2_{f} }{F^2}\,,
\label{eq:c1_I}  \\ 
\left(\frac{c_1}{F^4}\right)^{\rm min} 
&=& \frac{N_c y_t^2}{4\pi^2}   \frac{M_T^2}{F^2} 
-\frac{9 g_R^2}{32\pi^2}  \left(1 -\frac{g_R^2v^2_\Sigma }{2 M_\rho^2} \right) \frac{M_\rho^2}{F^2}
+\mathcal{O}(g_Lg_R,g_L^2)\,, \qquad
\label{eq:c1_II}
\eea
and of the $\sin^4(h/2F)$ term,
\be \label{eq:c2formula}
\frac{c_2}{F^4} =  \frac{N_c y_t^2}{4\pi^2} \frac{ M_T^2 }{F^2}  + \frac{9 g_R^2  }{32\pi^2}   \delta_{\pi}  
\left(1 - \frac{ g_R^2 v^2_\Sigma}{2 M_\rho^2} \right)
\frac{M_\rho^2}{F^2}
-\frac{9 g_R^4}{64\pi^2} \log\left(\frac{M_\rho^2}{M^2_{W_R}}\right)  +\mathcal{O}(g_Lg_R,g_L^2) \, .  
\ee
Let's start by analysing the result for $c_2$. The physical condition in \eqref{eq:c12phys} can be matched by keeping the fermion contribution only, which is expected to be parametrically larger than the gauge contributions here, in the limit of a light $M_T$. Note that this contribution is finite, hence it is largely insensitive to the details of the composite model but for the key assumption of light top partners.  More precisely, assuming $\delta_\pi \ll 1$ one needs $M_T \approx 2.5 F$, which is perfectly consistent with present bounds (see \S \ref{sec:pheno}). Higher $M_T$ values can be obtained at the expense of a modest tuning of $\delta_\pi$.  The term proportional to $g_R^4$ can be neglected if $g_R \lesssim 2$. 

Concerning $c_1$, the tuning needed to satisfy 
the physical condition in \eqref{eq:c12phys} requires two main ingredients. First, a cancellation of the leading (quadratically divergent) fermion contribution in \eqref{eq:c1_I}. 
This can be achieved imposing additional symmetries in the fermion mass terms forcing a cancellation between left-handed and right-handed 
contributions.\footnote{Extensive discussions of fermion representations and related symmetries that can be used to achieve this goal can be found in Refs.~\cite{Panico:2012uw,Csaki:2017cep}.}
Doing so, one is left with explaining the {\em minimal tuning}~\cite{Panico:2012uw} related to the term in \eqref{eq:c1_II}, where the fermion contribution is the same as in $c_2$ ({\em i.e.}~the model-independent contribution associated to the top-quark Yukawa coupling).
In our setup, a natural ingredient to achieve this tuning is a cancellation between the fermion and gauge contributions, which are predicted to have opposite sign. This cancellation does not happen with a flavour-universal gauge group since, in that case, the gauge coupling is too small. In order for the cancellation to take place in our setup, the following two conditions need to be satisfied
\begin{itemize}
\item{} A relatively large $g_R \equiv g_{R,3}$,  still within the perturbative regime, such that the gauge contribution reaches the same size as the fermion one. 
More precisely, we need $g_R^2 \times M^2_\rho/(6F)^2 \gtrsim 1$, that for natural values of $M_\rho$ implies $g_{R,3} =O(1) \gg g_{R,12} \approx g_Y^{\rm SM}$.  
\item{} A relatively light $M^2_{W_R}$, so as not to suppress (or even change the sign of) the gauge contribution.
As can be inferred from \eqref{eq:c1_II}, the condition required is 
\be
M^2_{W_R}  = \frac{1}{4} g_R^2 v^2_\Sigma < \frac{1}{2} M^2_\rho\,,
\ee
where the first equality will be derived explicitly in \S \ref{sec:GBmasses}. Note that to see the need for this second condition required carefully expanding the logarithm appearing in the Coleman--Weinberg potential to formally higher order before integrating, and relies on the fact that the terms proportional to $g_R^2$ {\em vs.} $g_R^4$ in \eqref{eq:c2formula} have strictly opposite sign.
\end{itemize}

\section{Phenomenology}
\label{sec:pheno}
The phenomenological implications of the model can be divided into two main categories: those related to the strong dynamics (such as modified Higgs-boson couplings, dynamics of the top partners and heavy resonances, {\em etc.})
and those related to the flavoured gauge bosons. The former category does not differ significantly from what is discussed in the composite Higgs model literature, hence we shall discuss this category of bounds rather briefly.

On the other hand, we present in more detail the phenomenology of the flavoured gauge bosons, which are specific to the flavour-deconstructed model.
As we have shown in the previous section, these exotic fields are predicted to be lighter than all composite states but for the Higgs boson and the top partners. 

\subsection{Constraints from the strong dynamics}

\subsubsection{Bounds on $F$ from Higgs couplings}
\label{sec:Fbounds}

We here derive the bound on the scale $F$ coming from modifications of the Higgs couplings to gauge bosons due to the strong dynamics.
The SM Higgs Lagrangian has the form
\be
\mathcal{L}_{H}^{\rm SM} = \left(D_\mu H\right)^\dagger \left(D^\mu H\right)-V_{\rm SM} (H)\,,  
\qquad D_\mu \equiv \partial_\mu -ig_L A_{L,\mu}^{a}\frac{\sigma_{a_L}}{2} 
-ig_Y B_\mu Y\,.
\ee
Here $H$ denotes the Higgs field in the doublet notation, such that 
\be
H_{\rm unit.-gauge}=\frac{1}{\sqrt{2}}
\begin{pmatrix} 0 \\ h^{\rm SM}_{\rm phys}(x)+v_{\rm EW} \end{pmatrix}\,,
\ee
where $v_{\rm EW} =(\sqrt{2}G_F)^{-1/2} \approx 246$ GeV.
From the kinetic term, the couplings of the physical Higgs boson to the $SU(2)_L$ SM gauge fields read
\be
g_{VVh}^{\rm SM}=g_L^2\frac{v_{\rm EW}}{4}\,,  \qquad g_{VVhh}^{\rm SM}=\frac{g_L^2}{8} \,.
\ee
In our composite model, the kinetic part of the Higgs Lagrangian is given in (\ref{kinU}). 
Expanding this Lagrangian in powers of $h(x)$ around its VEV  determined 
by the potential (\ref{eq:Vh})
leads to 
\be
\mathcal{L}_h \supset  g_L^2\frac{F^2}{2}\left[\sin^2\left(\frac{v}{2 F}\right)
+ \sin\left(\frac{v}{ F}\right)  \left(\frac{h}{2 F}\right)  
 + \cos\left(\frac{v}{ F}\right) \left(\frac{h}{2 F}\right)^2+\cdots\right] \left(A_\mu^i\right)^2 \, .
\ee
From the SM relation $m_W = \frac{1}{4}g_L^2 v_{\rm EW}^2$, we deduce
\be
v_{\rm EW}^2\equiv 4F^2\sin ^2 \frac{v}{2 F}\,.
\ee
Note that, consistently with the discussion in \S \ref{sec:GBL}, we recover $v\rightarrow v_{\rm EW}$
in the limit  $F\rightarrow \infty$.
Expressing $v$ in terms of $v_{\rm EW}$
we can relate the Higgs-gauge-boson  couplings of the composite model to the SM ones, obtaining 
\be
\label{higgscoupling}
g_{VVh}=g_{VVh}^{\rm SM}\sqrt{1-\xi}\,,\qquad g_{VVhh}=g_{VVhh}^{\rm SM}\left(1-2\xi \right)\,,
\ee
where, following~\cite{Contino:2010rs}, we have defined 
\be
\xi = \frac{v_{\rm EW}^2}{4F^2}
\equiv 
\frac{v_{\rm EW}^2}{\fHC^2}\,.
\ee
As expected, these results are perfectly consistent with those originally derived in~\cite{Agashe:2004rs} for the so-called minimal composite Higgs model.  

Up-to-date bounds from the LHC on the Higgs couplings can be found in~\cite{ATLAS:2020qdt,CMS:2020gsy}.
Combining ATLAS and CMS data leads to 
$g_{VVh}/g_{VVh}^{\rm SM} > 0.97~(95\%\,{\rm CL})$,
which implies 
 \be
\xi < 0.06~(95\%\,{\rm CL})
\qquad {\rm or} \qquad \fHC > 1.0~{\rm TeV}~(95\%\,{\rm CL})\,.
\label{eq:flimHVV}
\ee

\subsubsection{Top partners and heavier resonances} \label{sec:topp_pheno}

Light top partners are a general prediction of composite Higgs models 
and extensive searches have been performed at the LHC for them. A selection of 
recent results can be found in~\cite{ATLAS:2024gyc,CMS:2022fck,ATLAS:2023bfh,CMS:2024bni}.
The bounds based on pair production, 
which are the most stringent except in a few specific model-dependent constructions, 
imply 
\be
M_T \gtrsim  1.5 \mathrm{~TeV}\,. 
\ee
In order to  fulfil the relation 
$M_T \approx 2.5 F$ from the Higgs potential (see \S\ref{sect:HiggsPot}), this implies $\fHC \gtrsim 1.2$~TeV, delivering a slightly stronger limit  than the one set by the Higgs couplings in (\ref{eq:flimHVV}).

As far as the direct searches for heavy resonances are concerned, specifically on the vector resonances which are expected to be the lightest ones, present LHC limits on $W'$ and  $Z'$ bosons~\cite{ATLAS:2024qvg,CMS:2023gte} imply  $M_\rho \gtrsim 5$~TeV.
More stringent constraints are derived from indirect searches via electroweak observables. 
The  leading (tree-level) effects of heavy composite resonances in EW precision observables 
(such as the $S$ parameter or the $m_W/m_Z$ ratio) are of the order of
\be
\delta_{\rm EW} = g_{L,R}^2 \frac{v^2}{M^2_\rho}
\equiv \xi\frac{g_{L,R}^2 }{g^2_\rho}\,,
\ee
where we have defined $g_\rho \equiv  M_\rho/\fHC$.
Present bounds are satisfied for $\delta_{\rm EW} \lesssim 10^{-3}$,
which can be obtained  for a large enough $g_\rho$, well within the upper bound 
$g^{\rm max}_\rho \approx 4\pi$ set by perturbativity.

Aiming at the lowest possible value of $\fHC$ in order to minimize the tuning in the
Higgs potential, a reference benchmark point/region for the composite sector is provided 
by\footnote{The benchmark range for $g_\rho$ closely replicates a QCD-like spectrum.}
\begin{equation}
    \fHC \approx 1.5 \text{~TeV}, \qquad 
    M_T\approx 1.8\div 2.0 \text{~TeV}, \qquad
    g_\rho \approx 5\div 6
\end{equation}
%$\fHC \approx 1.5$~TeV, with  $M_T\approx 1.8\div 2.0$~TeV and  $g_\rho \approx 5\div 6$.
 This choice implies a $3\%$
 tuning in the Higgs potential ($\xi \approx 0.03$) with $O(1\%)$ corrections to the Higgs couplings. Corrections to the EW precision observables are slightly below the per-mille level, 
 and $M_\rho \approx 8\div 10$~TeV.

\subsection{Flavoured gauge bosons}
\label{sec:flavoredGB}

We now turn to the phenomenology associated with the flavour deconstruction, which gives rise to heavy gauge bosons. As described above, the flavour structure is generated thanks to the minimal symmetry breaking pattern
\begin{equation}
    U(1)_Y^\light \times U(1)_{B-L}^\heavy \times SU(2)_R^\heavy \to U(1)_Y,
\end{equation}
triggered by two link fields getting VEVs: $\Sigma \sim \left(\frac{1}{2},0,{\bf 2}\right)$,
which gets the VEV $\langle \Sigma \rangle = \frac{1}{\sqrt{2}}(0\,\,v_\Sigma)^T$, and $\Omega_q \sim \left(\frac{1}{6},-\frac{1}{6},{\bf 1}\right)$, which acquires the VEV $\langle \Omega_q \rangle = \frac{1}{\sqrt{2}}v_\Omega$.\footnote{For simplicity, unless otherwise stated, in this section we disregard the possibility of the additional link field $\Omega_\ell \sim \left(-\frac{1}{2},\frac{1}{2},{\bf 1}\right)$.}

\subsubsection{Gauge boson masses}
\label{sec:GBmasses}

To obtain the spectrum of the heavy gauge bosons, the relevant Lagrangian is   
\begin{align} 
    \cL \supset 
    \left|\partial_\mu \Sigma + i\frac{g_{R,3} \sigma^a}{2}  W_{a\mu}^\heavy\Sigma + i\frac{g_{Y,12}}{2} B^\light_\mu \Sigma \right|^2 +\left|\partial_\mu \Omega_q +  i\frac{g_{Y,12}}{6}B^\light_\mu \Omega_q - i\frac{g_{3,B-L} }{6}X^\heavy \Omega_q \right|^2
    \, , \nonumber
\end{align}
which contains the following mass terms after subbing in the VEVs:
\begin{align}\label{eq:Lscalar}
    \cL \supset \,\,  &\frac{1}{8}v_\Sigma^2 \left[ g_{R,3}^2(W_1^{\heavy\, 2} + W_2^{\heavy\, 2}) + (g_{Y,12} B^\light - g_{R,3} W_3^{\heavy})^2
    \right] \nonumber \\
    &\quad +\frac{1}{72}v_\Omega^2(g_{Y,12} B^\light - g_{B-L,3} X^{\heavy})^2
\end{align}
The three neutral gauge bosons all mix. The mass mixing matrix is
\begin{equation}
    \cL \supset \frac{1}{8}(B^\light\,\, X^\heavy\,\, W_3^\heavy)
    \begin{pmatrix}
        g_{Y,12}^2(v_\Sigma^2+\frac{1}{9}v_\Omega^2) & -\frac{1}{9}g_{Y,12} g_{B-L,3} v_\Omega^2 & -g_{Y,12} g_{R,3} v_\Sigma^2 \\
        -\frac{1}{9}g_{Y,12} g_{B-L,3} v_\Omega^2 & \frac{1}{9}g_{B-L,3}^2 v_\Omega^2 & 0 \\ -g_{Y,12} g_{R,3} v_\Sigma^2 & 0 & g_{R,3}^2 v_\Sigma^2
    \end{pmatrix}
    \begin{pmatrix}
        B^\light \\ X^\heavy \\ W_3^\heavy
    \end{pmatrix}\, .
\end{equation}
This matrix has zero determinant, so there remains a massless neutral gauge boson, corresponding to the unbroken SM $U(1)_Y$ subgroup. This massless field corresponds to the linear combination
\begin{equation}
    B \propto \,\, \frac{1}{g_{Y,12}} B^\light + \frac{1}{g_{B-L,3}} X^\heavy + \frac{1}{g_{R,3}}W_3^\heavy\,, \qquad m_B = 0\, ,
\end{equation}
which, sure enough, has flavour-universal couplings to the SM fermions (see below).
%The couplings of the massless eigenstate are, sure enough, the flavour-universal couplings of the SM $B$ boson: 

The masses of the two heavy neutral gauge bosons are quite complicated expressions of the couplings and the VEVs, in general. However, motivated by our considerations of the Higgs potential (plus collider phenomenology), we are most interested in the r\'egime
\begin{equation} \label{eq:Y12}
    g_{Y,12} \ll g_{R,3},\, g_{B-L,3}\, ,
\end{equation}
{\em i.e.}~with smaller gauge couplings to the light generations than to the third one. 
In this limit, the off-diagonal terms responsible for mixing are all suppressed, and one can determine the two heavy mass eigenvalues perturbatively in the small ratio of couplings:
%%%%%%%%%%%%%%%
    % \begin{equation}
    %     m_{Z_V}^2 \approx \frac{1}{4} g_{B-L,3}^2 v_\Omega^2 + \mathcal{O}(g_{Y,12}^2 v^2)\,, \qquad 
    %     m_{Z_R}^2 \approx \frac{1}{4} g_{R,3}^2 v_\Sigma^2 + \mathcal{O}(g_{Y,12}^2 v^2)\, .
    % \end{equation}
    % Actually, when we do precision EW phenomenology it will be important to keep the sub-leading mass corrections, because these change the $Z_R'$ and $W_R'$ mass ratio (upon which the $W$ mass shift depends). We have
\begin{equation}
    M_{Z_V}^2 \approx \frac{1}{36} (g_{B-L,3}^2 +g_{Y,12}^2) v_\Omega^2 + \mathcal{O}(g_{Y,12}^4)\,, \qquad 
    M_{Z_R}^2 \approx \frac{1}{4} (g_{R,3}^2 +g_{Y,12}^2) v_\Sigma^2 + \mathcal{O}(g_{Y,12}^4)\, .
\end{equation}

Finally, looking back to (\ref{eq:Lscalar}), there are massive eigenstates of the unbroken $U(1)_Y$ with charge $\pm 1$ (which can be read off from the non-abelian $SU(2)_R$ algebra, specifically the non-zero commutator between the unbroken $\sigma_3$ direction and the combinations $\sigma_1 \mp i\sigma_2$ corresponding to $W^\pm$):
 \begin{equation}
 W_R^{\pm} = \frac{1}{\sqrt{2}}(\tilde W_1 \mp i \tilde W_2)\,, \qquad M_{W_R}^2 = \frac{1}{4} g_{R,3}^2 v_\Sigma^2\,.
\end{equation}
To leading order, the mass ratio for the $Z_R$ and $W_R$ is
\begin{equation}
    \left(\frac{M_{Z_R}}{M_{W_R}}\right)^2 = \frac{g_{R,3}^2+g_{Y,12}^2}{g_{R,3}^2} \geq 1\, ,
\end{equation}
mirroring the formulae for the $Z$ to $W$ mass ratio in the SM.

\subsubsection{Gauge boson couplings to SM fields}

We now turn to the couplings of the mass eigenstate gauge bosons. We first derive the couplings to SM fermions, then to the Higgs.

\subsubsection*{Couplings to fermions}

Firstly, the massless gauge field $B$ couples to the flavour-universal SM hypercharge, 
$Y = Y^\light + (B-L)^\heavy + T_{R^3}^{\heavy}$, that is left unbroken by this symmetry breaking pattern irrespective of the values of the VEVs or the gauge couplings.\footnote{This reflects a general group theory statement that one only ever breaks to the diagonal subgroup, under certain conditions that are here satisfied -- see Refs.~\cite{Craig:2017cda,Davighi:2023xqn}.} 
The corresponding gauge coupling $g^\prime$ then satisfies the following matching condition
\begin{equation}
    \frac{1}{g^{\prime 2}} = \frac{1}{\gY^2}+\frac{1}{\gBL^2} + \frac{1}{\gR^2}\, .
\end{equation}
One can enforce this condition by parametrizing the UV gauge couplings via
\begin{equation}
    g'=\gY \cos\theta= \gBL \sin\theta\cos\phi = \gR \sin\theta \sin\phi \, .
\end{equation}
This let us trade the three couplings $\{\gY, \gBL, \gR\}$ for two polar angles $\{\theta,\phi\}$. The inverse coordinate transformation is
\begin{equation}
    \theta = \arctan\left(\frac{\gY}{\gBL\gR}\sqrt{\gBL^2+\gR^2} \right)\, , \qquad \phi = \arctan\left( \frac{\gBL}{\gR}\right)\, .
\end{equation}
Other useful relations are:
\begin{equation}
    \frac{\gY}{\gR} = \tan\theta \sin\phi, \qquad 
    \frac{\gY}{\gBL} = \tan\theta \cos\phi,
\end{equation}
In these coordinates, the limit $\gY\ll \gR,\, \gBL$ that we are interested in translates to the small angle limit
\begin{equation}
\theta \ll 1, \qquad \phi \sim 1\, .    
\end{equation}

The two heavy mass eigenstates coincide with the gauge eigenstates up to small corrections: we have $Z_{V} \approx X^\heavy - \frac{g_{Y,12}}{g_{B-L,3}} B^\light + \mathcal{O}(g_Y^2) W_3^\heavy$ and
$Z_R \approx W_3^\heavy - \frac{g_{Y,12}}{g_{R,3}} B^\light + \mathcal{O}(g_Y^2) X^\heavy$.
The notation $Z_V$ and $Z_R$ reminds us that the former neutral boson couples nearly vector-like to fermions, while the latter is nearly right-handed.
To deduce the couplings to SM fields, we invert these relations: 
\begin{align}
    B^\light &\approx B - t_\theta c_\phi Z_V  - t_\theta s_\phi Z_R  \, , \\
    X^\heavy &\approx Z_V  + t_\theta c_\phi B\, , \\
    W_3^\heavy &\approx Z_R  + t_\theta s_\phi B\, ,
\end{align}
where $s_\theta:=\sin\theta$ {\em etc.}
We can thence deduce the couplings of both $Z^\prime$ bosons to SM fermions, which take the form
\begin{equation}
    \cL \supset Z_{V \mu} 
    J_{V,\psi}^\mu + Z_{R \mu}J_{R,\psi}^\mu\, ,
\end{equation}
where the relevant fermionic currents are found to be
\begin{align}
    J_{V,\psi}^\mu &= g_{B-L,3} J_{B-L}^{\heavy \mu} - \frac{g_{Y,12}^2}{\gBL} J_Y^{\light \mu}  = \frac{g^\prime}{s_\theta c_\phi} \left(J_{B-L}^{\heavy \mu} - t_\theta^2 c_\phi^2 J_Y^{\light \mu}  \right)\, ,
    \\
    J_{R,\psi}^\mu &= g_{R,3}J_R^{\heavy \mu} - \frac{g_{Y,12}^2}{\gR} J_Y^{\light \mu}  = \frac{g^\prime}{s_\theta s_\phi} \left(J_{R}^{\heavy \mu} - t_\theta^2 s_\phi^2 J_Y^{\light \mu}  \right)\, .
\end{align}
We notice from these formulae that one cannot take the coupling to light generations to zero ($\theta \to 0$) without the coupling to the third family simultaneously diverging.\footnote{There is also, in principle, a small deviation in the light family couplings due to the mixing between SM fermions and the heavy vector-like fermions $F_{R,L}^q$ that are integrated out at the high scale to generate the Yukawa hierarchies. The effects of such mixing has been taken into account in previous phenomenological studies of flavour deconstruction {\em e.g.}~\cite{Cornella:2021sby}, but for our purposes it is good enough to ignore these small mixing effects. Their inclusion would change the high-$p_T$ bounds derived below by a small amount. } 
This is the case for any flavour deconstruction setup: one cannot totally decouple the heavy gauge bosons from the light SM generations.

Finally, there is the coupling of the charged current gauge bosons $W_R^{\pm}$ that come from breaking the off-diagonal generators in $SU(2)_R^\heavy$. This gives a charged current interaction with third generation right-handed quarks:
\begin{equation} \label{eq:Wp_quark_current}
    \cL \supset - \frac{1}{\sqrt{2}}  \gR W_{R  \mu}^{-} \overline{d}_R^3  \gamma^\mu u_R^3 + \mathrm{h.c.}
\end{equation}
There is no leptonic coupling because, even if there are right-handed neutrinos, they are assumed heavy and already integrated out. Note this also decouples the high-$p_T$ bound because we have no semi-leptonic operators coming from integrating out the $W_R$ that modify $\ell\ell$ or $\ell\nu$ final states.

\subsubsection*{Couplings to the Higgs}

The massive gauge bosons generated by the $SU(2)_R^\heavy$ breaking couple also to the Higgs fields. To derive these couplings, it is convenient to use the bi-doublet notation $\Phi$ for the Higgs field that we introduced above in Eq.~(\ref{eq:Hdefinition}). This is related to the Higgs $SU(2)_L$ complex doublet $H$ via
\begin{equation}
    \PP = \begin{pmatrix}
        \hh^c & \hh
    \end{pmatrix}, \qquad  
    \hh^c = i\sigma_2 \hh^\ast\, .
\end{equation}
The covariant derivative acting on $\PP$
was given above in Eq. (\ref{eq:cov-deriv-Phi}), by which $SU(2)_L$ ($SU(2)_R^\heavy$) gauge bosons act straightforwardly by left (right) matrix multiplication.

To obtain the linear couplings of the heavy gauge fields to the Higgs doublet $\hh$ (which is all we need for tree-level matching to the SMEFT at dimension-6), we expand the kinetic part of the action $\cL = \frac{1}{4} \mathrm{Tr} \left[D_\mu \PP^\dagger D_{\mu} \PP \right]$, and isolate the piece that is linear in the $A_R$ gauge field. This current is
\begin{equation} \label{eq:LHiggsGauge}
    L \supset \frac{1}{4}\mathrm{Tr} \left[i g_{R,3} \partial_\mu \PP^\dagger \left(\PP.\frac{\sigma^a}{2} A_{R\mu}^a \right) -ig_{R,3}\left( \frac{\sigma^a}{2} A_{R\mu}^a .\PP^\dagger\right)  \partial_\mu \PP  \right]\,.
\end{equation}
After the symmetry breaking described above, we can deduce the couplings to the heavy charged current bosons by rotating to the $(W_R^{\pm}, Z_R \approx A_R^3)$ basis. The charged-current interaction then assumes the form
\begin{align}
    L \supset &\frac{g_{R,3}}{2\sqrt{2}} W^-_{R \mu} \, \bigg[-\phi_1 \partial_\mu \phi_4+\phi_2 \partial_\mu \phi_3 - \phi_3 \partial_\mu \phi_2 + \phi_4 \partial_\mu \phi_1  \\
    &\qquad \qquad + i(\phi_1\partial_\mu \phi_3 +\phi_2 \partial_\mu \phi_4 - \phi_3 \partial_\mu \phi_1 - \phi_4 \partial_\mu \phi_2 )\bigg]  + \mathrm{h.c.} \nonumber \\
    =&-\frac{g_{R,3}}{2\sqrt{2}}W_{R\mu}^- i D_\mu \hh^T i\sigma_2 \hh + \mathrm{h.c.}~  
\end{align}
while the neutral-current coupling is
\begin{align}
    L \supset\,\, &\frac{1}{2}\gR Z_R^{\mu} \left( \phi_1 \partial_\mu\phi_2 - \phi_2 \partial_\mu \phi_1 - \phi_3 \partial_\mu \phi_4 + \phi_4 \partial_\mu \phi_3  \right) \\
    =&\,\, \frac{\gR}{4}Z_R^{\mu} (i \hh^\dagger D_\mu \hh + \mathrm{h.c.}) 
    ~ \equiv~  Z_R^{\mu} J_{R,H\, \mu} \, ,
\end{align}
 where the last equality defines $J_{R,H}^\mu$.

\subsubsection{Matching to the SMEFT} \label{sec:SMEFT}

Having derived the couplings of the heavy vector bosons, we can now integrate out at tree-level and match to the dimension-6 SMEFT to estimate the most important phenomenological consequences. We do so using the tree-level matching dictionary of Ref.~\cite{deBlas:2017xtg}.

When the two neutral gauge bosons are integrated out, we can write down the SMEFT Lagrangian that results in terms of the currents derived above:
\begin{align}
    \cL \supset -\frac{J_{V,\psi}^\mu J^{\phantom\mu}_{V,\psi\, \mu}}{2M_{Z_V}^2} -\frac{J_{R,\psi}^\mu J^{\phantom\mu}_{R,\psi\, \mu}}{2M_{Z_R}^2} 
    -\frac{J_{R,\psi}^\mu J^{\phantom\mu}_{R,H\, \mu}}{M_{Z_R}^2} -\frac{J_{R,H}^\mu J^{\phantom\mu}_{R,H\, \mu}}{2M_{Z_R}^2}
\end{align}
The first term comes from the $Z_V$ boson, which couples only to fermions and hence produces only 4-fermion operators. The latter three terms come from integrating out the $Z_R$ boson, containing 4-fermion operators with various chiral structures (second term), mixed fermion-boson operators of the form $\mathcal{O}_{Hf}\sim (\hh^\dagger i \overleftrightarrow{D}_\mu \hh)(\overline{f}\gamma^\mu f)$ (third term), and pure bosonic operators $\mathcal{O}_{HD}\sim (\hh^\dagger D_\mu \hh)^2$ and $\mathcal{O}_{H\Box}\sim |\hh|^2 \Box |\hh|^2$ (fourth term). This accounts for all tree-level dimension-6 operators produced by integrating out the two heavy neutral gauge bosons. All these operators will lead to important experimental constraints, which we discuss in the following subsections.

Integrating out the charged current $W_R^{\pm}$ bosons gives additional operators. Starting with four-fermion operators, only the SMEFT operators  $\mathcal{O}_{ud}^{(1)}$ and $\mathcal{O}_{ud}^{(8)}$ are generated given the coupling (\ref{eq:Wp_quark_current}), with suppression factors for all light-family flavour components. These are subject only to very weak experimental bounds, and will play no role in the phenomenological analysis that follows. Concerning mixed fermion-boson operators, different structures are generated by the charged current (compared to neutral one). Firstly, we get modified Yukawa operators $\mathcal{O}_{eH} \sim |H|^2 \overline{l} H e$, $\mathcal{O}_{dH}$ and $\mathcal{O}_{uH}$ (defined similarly). 
%while the bounds are of order a few TeV from $H\to \tau\tau$ and $H \to bb$ signal strengths for order-1 WCs, here 
These operators are generated only via the Higgs interaction, and so their Wilson coefficients are suppressed by SM Yukawa couplings:
	\begin{equation}
		C_{eH,\, dH,\, uH} \sim y_{\tau,\, b,\, t}\, \frac{\gR^2}{16M_{W_R}^2}\,.
  \label{eq:modY}
	\end{equation}
These operators modify the $H\to \tau\tau$ and $H \to bb$ signal strengths; however, the suppression by $y_\tau$ ($y_b$) in (\ref{eq:modY}) implies  
the corresponding bounds are weak. Integrating out the $W_R$ also generates the operator $\mathcal{O}_{Hud}^{ij} \sim (H^\dagger iD_{\mu} H) \overline{u}_R^i \gamma^\mu d_R^j$, with unsuppressed Wilson coefficient for the $i=j=3$ component: this operator has a significant impact in flavour physics (see \S \ref{sec:flavour_pheno}). Lastly, the $W_R$ also gives contributions to pure bosonic operators, in particular an additional contribution to $\mathcal{O}_{HD}$, which modifies the 
SM prediction for $m_W$ (\S \ref{sec:EW_pheno}). Both these effects are discussed in detail below.

\subsubsection{Flavour observables} \label{sec:flavour_pheno}

The heavy gauge bosons in this theory have flavour non-universal couplings to the SM fermions, but respect an accidental $U(2)^5$ global flavour symmetries acting on the light families, 
 with a minimal breaking structure linked to the subleading entries in the Yukawa couplings. This symmetry allow them to be close to the TeV scale without contravening the tight flavour bounds for flavour-changing processes in the 1-2 sector (such as those from kaon mixing). 
 
 Nonetheless, there are important bounds coming from flavour-changing processes involving bottom quarks, which we here survey:
\begin{itemize}
    	\item {\bf Bound on $W_R$ mass from $B \to X_s \gamma$.} As anticipated, the   $W_R$ boson generates a tree-level contribution to  $O_{Hud} \sim (H^\dagger iD_{\mu} H) \overline{u}_R \gamma^\mu d_R$, with unsuppressed coupling
     for third generation quarks,
	\begin{equation}
		C_{Htb} = -\frac{g_{W_R}^H g_{W_R}^{b_R t_R}}{M_{W_R}^2} = \frac{1}{v_\Sigma^2}\, .
	\end{equation}
    This operator is severely constrained by its one-loop contribution to $B \to X_s \gamma$.	Using the results of~\cite{Allwicher:2023shc}, we deduce 
 %\mar{Should be 3.2 TeV, table 14 from Lukas}
    \begin{equation}
        v_\Sigma \geq 3.2 \text{~TeV}\,.
    \end{equation}
    Note that this effect is independent of the flavour orientation of the Yukawa couplings, since the flavour mixing come from the CKM matrix (or the flavour-off-diagonal left-handed coupling of the $W$
    field in the mass-eigenstate basis). The diagonalization of the Yukawa couplings induce, 
    in principle, also non-vanishing contributions to the $O_{Hud}$ operator with light quarks. However, the corresponding couplings are very suppressed 
     by the negligible amount of flavour mixing in the right-handed sector, hence can be safely neglected. 
    \item 
    {\bf Bounds on $Z^\prime$ masses  from FCNC processes.} The smallness of flavour mixing in the right-handed sector implies the $Z_R$ boson is weakly constrained  from FCNC processes. The situation is different for the $Z_V$ boson, where left-handed flavour mixing (from the diagonalisation of the Yukawa couplings) can induce sizeable 
    FCNC couplings.  The corresponding bounds on $v_\Omega$,  dominated by $B_s$-mixing, can be deduced from the recent analysis in Ref.~\cite{Barbieri:2023qpf}. They reach 
    \be
    v_\Omega \gtrsim 2.7 \mathrm{~TeV }
    \ee
    in the case of pure up-alignment, but could be reduced with respect to this value by assuming some degree of down-alignment. 
    \item {\bf Flavour constraints on the composite states.}
    On general grounds, the composite states (top-partners, vector resonances, and higher-mass states) also mediate flavour-changing interactions. First, it is worth noting that
    the minimal breaking of the flavour symmetry prevents tree-level contributions to FCNCs suppressed only by the light top partner mass. As far as other effects are concerned,
    an extensive analysis in various classes of composite models has been presented recently in Ref.~\cite{Glioti:2024hye}. The case there denoted `partial Right Universality' (pRU) is quite close to our framework in this respect, since our deconstructed gauge symmetry structure enshrines the pRU global symmetry pattern as accidental. From the analysis of Ref.~\cite{Glioti:2024hye} we deduce that the bounds from flavour observables are all satisfied if $g_\rho$ is large enough to satisfy the EW bounds (see \S \ref{sec:topp_pheno}). 
    
\end{itemize}

\subsubsection{Higgs and electroweak observables} \label{sec:EW_pheno}

The heavy gauge bosons coming from the $SU(2)_R^\heavy$ breaking couple to the Higgs, giving rise to mixed fermion-Higgs operators and pure Higgs operators at tree-level (see \S \ref{sec:SMEFT}). These operators give sizeable effects in electroweak precision observables (EWPO), principally the $Z$-pole measurements made at LEP-II as well as measurements of the $W$ mass. We here pinpoint the most important observables and estimate the strength of constraints to guide future studies. We do not undertake a comprehensive fit to data in this work. 

\subsubsection*{$W$-boson mass}

The heavy $W^\pm_R$ and $Z_R$ bosons leads to a shift ($\delta m_W$) 
in the prediction for the $W$ mass. The leading effect is 
related to their tree-level contribution to the  Wilson coefficient $C_{HD}$:\footnote{In principle, there is also a contribution from the four-lepton operator $C_{ll}^{1221}$, induced by integrating out the $Z^\prime$ bosons, which arises because this operator modifies the extraction of $G_F$ from the muon decay. However, this operator is lepton-flavour-violating and is induced by a rotation in the lepton 1-2 sector that must be sufficiently small to satisfy other constraints from $\mu \to 3e$ and $\mu \to e\gamma$. As a result, this contribution to $\delta m_W$ turns out to be safely  negligible.  } 
\begin{equation}
 C_{HD} = \left(\frac{|g_{W_R}^H|^2}{M_{W_R}^2} -\frac{2|g_{Z_R}^H|^2}{M_{Z_R}^2}\right) = \frac{\gR^2}{8} \left(\frac{1}{M_{W_R}^2} -\frac{1}{M_{Z_R}^2}   \right)\, .
 \label{eq:CHD}
\end{equation}
Notice that (\ref{eq:CHD}) vanishes in the `custodial' limit $M_{W_R} = M_{Z_R}$. This would occur in principle for pure $SU(2)_R^{\heavy} \to \{0\}$ breaking, i.e.~without the linking which results in the $Z_{R}$ mixing with the light-family-hypercharge boson 
(that shifts $M_{Z_R}$ without affecing $M_{W_R}$). 
Accounting for this mixing, we have the following non-zero $C_{HD}$ at tree-level
\begin{equation}
    C_{HD} = \frac{1}{2v_\Sigma^2}\left[ 1 - \left(1+\frac{\gY^2}{\gR^2}\right)^{-1}\right] \approx \frac{1}{2v_\Sigma^2}\frac{\gY^2}{\gR^2}\, .
\end{equation}
To a good accuracy, the shift in the $W$-mass is given by~\cite{Bjorn:2016zlr} $\delta m_W^2/{m_W^2} \approx \frac{1-s_w^2}{1-2s_w^2} C_{HD} v^2/2$,
where $\delta m_W^2 := (m_W^2)_{\text{SM}} - (m_W^2)_{\text{SMEFT}}$ and $s_w$ is the sine of the Weinberg angle. Using also $\delta m_W^2/m_W^2 \approx 2 \delta m_W/m_W$, we find
\begin{equation} \label{eq:Wmass}
    \delta m_W \approx \frac{1-s_w^2}{1-2s_w^2}  \frac{v_{\mathrm{EW}}^2}{8v_\Sigma^2}\frac{\gY^2}{\gR^2}\, m_W =\left( \frac{9\text{~TeV}}{v_\Sigma}\right)^2 \frac{\gY^2}{\gR^2} \times (10 \text{~MeV})\, .
\end{equation}
As expected, the shift in the $W$ mass vanishes in the limit $\theta \to 0$ (due to the custodial protection). Hence the effect is small in the limit where the exotic gauge bosons couple dominantly to the third family -- the same limit that is also strongly favoured by collider searches (see below), {\em as well as} by our independent consideration of tuning in the Higgs potential.
Importantly, the shift in the $W$ mass is necessarily positive (with respect to the SM prediction), going in the same direction as a current small deviation in combined measurements of the $W$ mass~\cite{LHC-TeVMWWorkingGroup:2023zkn}.
The size of this deviation (roughly the same size as the current uncertainty on the measurement) is a few tens of MeV.
From (\ref{eq:Wmass}) we see that, taking as a benchmark $v_\Sigma = 3$ TeV, one has $\delta m_W \approx 10$ MeV for $\gY \approx \gR/3 \implies \gR \approx 1$, exactly in the region favoured by our considerations on the Higgs potential in \S \ref{sect:HiggsPot}.

We remind the reader that we have in mind a region of parameter space in which the effects of the composite sector on EWPOs are at the per-mille level (see \S \ref{sec:topp_pheno}), giving just small corrections to the leading-order effects from the (lighter) $Z_R$ and $W_R$ gauge bosons that we consider here.

\subsubsection*{$Z$-pole observables}
Other EWPOs involve modifications to the electroweak gauge boson couplings to SM fermions, induced from integrating out the $Z_{R}$ gauge boson (that mixes with the $Z$). Here the leading effects are due to the mixed fermion-Higgs operators $\mathcal{O}_{Hf}$. 

Because of the right-handed chiral structure of the coupling to third generation fermions, we have  $C_{Hl}^{33} = C_{Hq}^{(1)\, 33} =0$.  Wilson coefficients 
involving light-generation fermion currents (which are proportional to the hypercharge of the fermion field involved) are suppressed by the small mixing parameter $\sin\theta$:\footnote{Note that we can further reduce EWPO effects involving light leptons and quarks by considering the $\phi \ll 1$ (i.e $\gR \ll \gBL$) limit.
}
\begin{equation}
    \{C_{Hl}^{11,22}, C_{He}^{11,22}, C_{Hq}^{(1)\,11,22}, C_{Hu}^{11,22}, C_{Hd}^{11,22} \} \approx \frac{s_\theta^2 s_\phi^2}{6 v_\Sigma^2} \{-3,-6, +1, +4, -2 \}
\end{equation}
The largest effects, however, involve the right-handed third family fermion fields. These come from the unsuppressed Higgs-bifermion operators:
\begin{equation}
    C_{He}^{33} \approx -C_{Hu}^{33} \approx C_{Hd}^{33} \approx \frac{1}{2v_\Sigma^2}\, .
\end{equation}
We can estimate the strength of the bound from EWPOs by taking the one-at-a-time constraints on these three coefficients. 
The analysis of Ref.~\cite{Allwicher:2023shc} suggests the strongest of such bound comes from the  modified $Z$ coupling to right-handed taus, thanks to the LEP measurement of  $R_\tau = \Gamma(Z \to \tau^+ \tau^-)/\Gamma(Z \to q \bar q)$. From the anlsysis of 
Ref.~\cite{Allwicher:2023shc} we deduce 
\begin{equation}
    v_\Sigma \gtrsim 2.7 \text{~TeV}\, ,
\end{equation}
which is comparable to the $B\to X_s \gamma$ bound discussed in \S \ref{sec:flavour_pheno}. We therefore identify these kind of precision flavour and electroweak observables involving third generation fermions (taus and bottoms) as particularly promising probes of a flavour deconstructed composite Higgs scenario.

\subsubsection{LHC bounds from Drell-Yan data}

Lastly, bounds on the $Z^\prime$ bosons masses can be derived from $pp$ collisions at the LHC.
We present here an EFT estimate of such bounds, as derived from high-$p_T$ Drell--Yan production of di-lepton pairs, measured by ATLAS~\cite{ATLAS:2020zms} and CMS~\cite{CMS:2021ctt}. These analyses use a total integrated luminosity of 139 fb$^{-1}$ (140 fb$^{-1}$) respectively. 

While a light enough $Z^\prime$ in the $s$-channel would generate a peak in the $m_{\ell\ell}$ distribution, we here approximate the impact of heavy $Z^\prime$ bosons by the four-fermion contact interaction the induce. As pointed out in various studies (see {\em e.g.}~\cite{Greljo:2017vvb}),   the inclusion of four-fermion operators  lifts the tail of the $m_{\ell\ell}$  distribution allowing to derive stringent bounds also for $Z^\prime$ masses well above the kinematical threshold. We use the \texttt{HighPT} package~\cite{Allwicher:2022gkm,Allwicher:2022mcg} in `EFT mode' to approximate a $\chi^2$ likelihood function for our model, turning on all dimension-6 semi-leptonic operators obtained by tree-level matching of the two heavy $Z^\prime$ bosons to the SMEFT. We include both light lepton and tau channels. 

\begin{figure}
    \centering
    \includegraphics[width=0.45\textwidth]{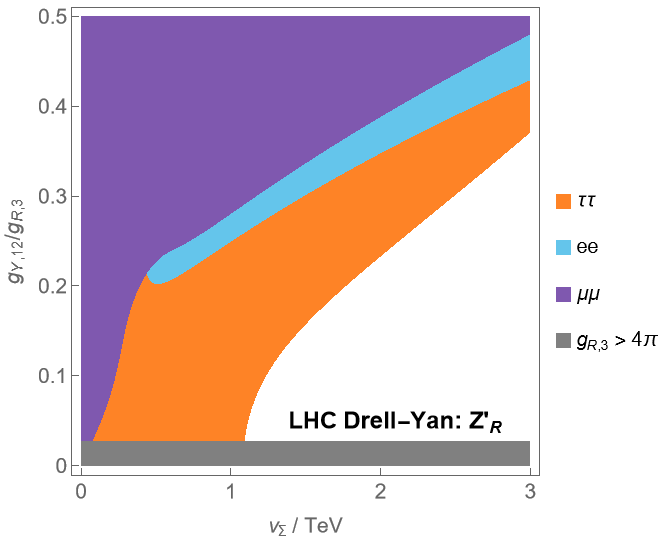}~~~~\includegraphics[width=0.46\textwidth]{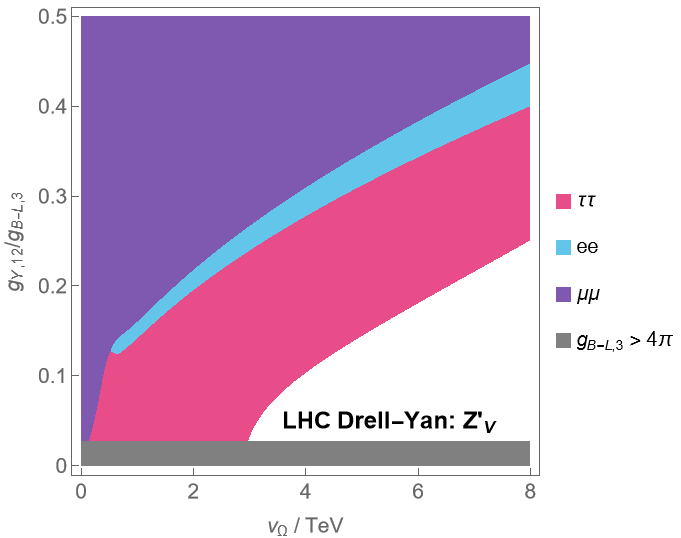}
    \caption{Constraints from high-$p_T$ Drell--Yan data at the LHC on the two $Z^\prime$ bosons arising from the flavour deconstruction in our model. In the region of parameter space plotted, for which $\gY\ll\gR,\, \gBL$, the constraint from $pp\to \tau\tau$ is the strongest. 
    }
    \label{fig:highPT1}
\end{figure}

The results are shown in Fig.~\ref{fig:highPT1}.
We present the constraints in terms of the relevant VEV $v_\Sigma$ ($v_\Omega$), vs.~the ratio of the $\gY$ to $\gR$ ($\gBL$) gauge coupling, which is assumed to be small. As can be seen, the high-$p_T$ constraints on the $Z_R$ boson even permits
$v_\Sigma \gtrsim 1 \text{~TeV}$,
in the limit $\gY \ll \gR$ ($\theta \ll 1$), when $\gR$ approaches $4\pi$. 
For more realistic  $\gR$ values, namely for $\gR \lesssim 1.5$, we get 
\be
v_\Sigma \gtrsim 2.0 \text{~TeV}\,.
\ee
It is the $pp \to \tau \tau$ search that set the most stringent limits the small $\theta$ region. Looking at Fig.~\ref{fig:highPT1}, one can see that the r\'egime of light $v_\Sigma$ (compared to the mass $M_\rho$ of the lightest spin-1 composite resonance) favoured by our considerations of the Higgs potential is, for now, completely consistent with high-$p_T$ LHC data. Once more, this is because of the flavour non-universal structure of the $Z_R$ couplings, with suppressed interactions with the light quarks. From our previous considerations, we expect the indirect tests (from both flavour, in particular $B \to X_s \gamma$, and electroweak, in particular the $R_\tau$ ratio) to be slightly more relevant than high-$p_T$ bounds at present; which may of course change as the LHC program advances.  

The second $Z_V$ boson, that couples mostly to the $J_{B-L}^\heavy$ fermion current, has a seemingly stronger bound on its VEV, $v_\Omega \gtrsim 3 \div 6$ TeV or so depending on the desired upper limit for the coupling. This factor of roughly 3 difference in the bounds just comes from the fact that the gauge boson mass is related to the VEV as $M_{Z_V} \approx \frac{1}{6}\gBL v_\Omega$, vs.~$M_{Z_R} \approx\frac{1}{2}\gR v_\Sigma$,
in units where the fermion charges are comparable. Actually, the relation between 
 $M_{Z_V}$ and $v_\Omega$ changes to $M_{Z_V} \approx \frac{1}{2}\gBL v_\Omega$ if the extra link field $\Omega_\ell$ (see Table~\ref{tab:Matter_Content}) is taken into consideration. In that case the bounds on 
 $v_\Omega$ and $v_\Sigma$ turn out to be very comparable.

We further recall that the scale $v_\Omega$ plays no role in the Higgs potential because the Higgs does not couple to this (dominantly vector-like) gauge boson; thus, the constraint on $v_\Omega$ does not have implications for naturalness, in contrast to $v_\Sigma$ which is crucially tied to the generation of the electroweak scale within our model.

\section{Conclusions and Outlook}
\label{sec::conclusion}

In this paper we have explored a model  for addressing the origin of flavour and the stability of the electroweak scale, by combining the idea of flavour non-universality with Higgs compositeness. In the UV, flavour non-universal gauge interactions encode the $2+1$ family structure, motivated by the observed pattern in the Yukawa couplings and the smallness of flavor-violating effects involving the light families. On the other hand, the Higgs mass is protected from quantum corrections induced by heavier degrees of freedom thanks to its  compositeness. The Higgs emerges as a light pNGB of a spontaneously broken symmetry and is unaffected by dynamics above the compositeness scale -- in particular, it is insensitive to the mass scale of the heavy fermion which is responsible for generating the hierarchical flavour structure needed to solve the flavour puzzle. 

The Higgs potential is generated at one-loop by explicit symmetry-breaking terms in the fermion sector and by the gauging of a subgroup of the strong symmetry. We have quantified the fermionic, scalar, and gauge contributions to the Higgs mass. We found that the flavour non-universal gauge structure of our model combined with a close-by symmetry breaking scale allows for a cancellation between fermionic and gauge contributions in the Higgs potential, thereby justifying {\em a posteriori} the observed little hierarchy between  
$v$ and $\fHC$. This important feature is connected to the possibility of having a large enough gauge coupling $g_{R, 3}=O(1) \gg g_{R, 12} \approx g_Y^{\mathrm{SM}}$, together with relatively light massive flavoured gauge bosons, $M_{W_R} \sim $~few TeV. 
It is important to stress that this finding is  reasonably general and would hold also in variations of the model, especially as far as the UV gauge group is concerned. The key observation is that combining partial compositeness with flavour deconstruction results in a more predictive framework than when each hypothesis is considered independently.

The interesting interplay between the scale of flavour deconstruction and the tuning in the Higgs potential motivated us to perform a study of the phenomenological implications of the model, focusing in particular to the non-standard effects induced by the flavoured gauge bosons.
 The masses and couplings of the latter turn out to be mostly constrained by $B \rightarrow X_s \gamma$ and $Z$-pole observables and, to a lesser extent (at least at present), by LHC measurements of $\sigma(pp\to \tau^+\tau^-)$. Present data allow these bosons to have masses in the vicinity of the TeV scale, as required by the consistency of the model. Interestingly enough, signals of their presence could appear via a modified high-$p_T$ behaviour of $pp\to \tau^+\tau^-$ in the high-luminosity phase of the LHC.

 The non-standard effects induced by the flavoured gauge bosons
are a distinct feature of the  flavour-deconstruction hypotheses and, to large extent, are independent of the strong dynamics originating the Higgs boson. These effects should be accompanied by other non-standard phenomena, such as modified Higgs-boson couplings and the presence of light top partners, which are related to the strong dynamics. The manifestation of both sets of phenomena is what could allow a clear experimental identification of the proposed model. The whole `natural' parameter space of the model could be probed  via flavour and electroweak observables, and modified Higgs couplings,  at a future $e^+e^-$ collider running both at the $Z$ pole and above the $Zh$  threshold.\footnote{Studies of the FCC-ee potential for the electroweak observables, and corresponding physics reach for (part of) the flavoured gauge bosons, can be found in Ref.~\cite{Allwicher:2023shc,Davighi:2023evx,Davighi:2023xqn,Stefanek:2024kds}}

On the theory side, one direction forward is to further speculate regarding the next layer of UV physics underlying the framework we have presented. Quite conspicuously, the model we sketched features fundamental scalar link fields $\Sigma_R$ and $\Omega$ which would themselves introduce (large) hierarchy problems.
One approach is to realise the scalar link fields also as composite particles of an enlarged strong sector. For instance, one can embed $\Sigma_R$ alongside the Higgs as composite pNGBs arising from a strong sector with global symmetry breaking
\begin{equation}
Sp(6)_{\mathrm{global}} \longrightarrow SU(2)_L \times SU(2)_R^\heavy \times SU(2)_R^\light    \, .
\end{equation}
This delivers pNGBs transforming in bi-fundamentals of each pair of $SU(2)$s, corresponding to the Higgs pNGB, the $\Sigma_R$, plus an extra bidoublet charged under $SU(2)_L \times SU(2)_R^\light$ (which would need to have a vanishing VEV). The natural mass scale for these pNGBs is around $f_{\mathrm{HC}}\sim \mathrm{TeV}\sim v_\Sigma$, so the $\Sigma_R$ state is completely naturally described as a composite pNGB; it is only the pNGB corresponding to the SM Higgs that must end up anomalously light due to fine-tuning in the loop-generated potential.

\acknowledgments 

We are grateful to Lukas Allwicher for helpful discussions.
This work is funded by the European Research Council (ERC) under the European Union’s Horizon 2020 research and innovation programme under grant agreement 833280 (FLAY), and by the Swiss National Science Foundation (SNF) under contract 200020-204428.

%\newpage
\appendix
\section{Further details on the CCWZ formalism}
\label{sec:Gamma}
The decomposition of $U^{\dagger} \partial_\mu U$ in \eqref{uderu} can be ordered in terms of the number of $\phi$-insertions. The LO and NLO terms in $\Gamma_\mu^a$ and $u_\mu^a$ are given by:
\begin{align}
\label{gammanu}
&\Gamma_\mu^a = -\frac{1}{2} \phi_b\partial_\mu\phi_c \cdot f_U^{cba} +\frac{1}{6}\phi_b \phi_c \partial_\mu \phi_d  \cdot f_B^{cde}f_U^{bea} + \ldots \\
&u_\mu^a = \partial_\mu \phi _a -\frac{1}{2} \phi_b\partial_\mu\phi_c  \cdot f_B^{cba} +\frac{1}{6}\phi_b \phi_c \partial_\mu \phi_d  \cdot\left(f_B^{cde} f_B^{bea}-f_U^{cde} f_{UB}^{eba}\right) + \ldots
\end{align}
where the commutation relations among unbroken generators $T_U^a$ and broken generators $T_B^a$ are given by:
\begin{align}
    &\left[T_B^a,T_B^b\right]=if_B^{abc}\cdot T_B^c+if_U^{abc}\cdot T_U^c \\
    &\left[T_U^a,T_B^b\right]=if_{UB}^{abc}\cdot T_B^c \, .
\end{align}
Notice that the LO contribution to $\Gamma_\mu^a$ starts at second order in the $\phi_a$, and hence plays no role in the expansion of $\mathcal{L}_U^{(2)}$ up to quadratic terms in $H$.

\bibliographystyle{JHEP}
\bibliography{refs}

\providecommand{\href}[2]{#2}\begingroup\raggedright\begin{thebibliography}{10}

\bibitem{DAmbrosio:2002vsn}
G.~D'Ambrosio, G.~F. Giudice, G.~Isidori and A.~Strumia, \emph{{Minimal flavor
  violation: An Effective field theory approach}},
  \href{https://doi.org/10.1016/S0550-3213(02)00836-2}{\emph{Nucl. Phys. B}
  {\bfseries 645} (2002) 155--187},
  [\href{https://arxiv.org/abs/hep-ph/0207036}{{\ttfamily hep-ph/0207036}}].

\bibitem{Davighi:2023iks}
J.~Davighi and G.~Isidori, \emph{{Non-universal gauge interactions addressing
  the inescapable link between Higgs and flavour}},
  \href{https://doi.org/10.1007/JHEP07(2023)147}{\emph{JHEP} {\bfseries 07}
  (2023) 147}, [\href{https://arxiv.org/abs/2303.01520}{{\ttfamily
  2303.01520}}].

\bibitem{Allwicher:2023shc}
L.~Allwicher, C.~Cornella, G.~Isidori and B.~A. Stefanek, \emph{{New physics in
  the third generation. A comprehensive SMEFT analysis and future prospects}},
  \href{https://doi.org/10.1007/JHEP03(2024)049}{\emph{JHEP} {\bfseries 03}
  (2024) 049}, [\href{https://arxiv.org/abs/2311.00020}{{\ttfamily
  2311.00020}}].

\bibitem{Contino:2003ve}
R.~Contino, Y.~Nomura and A.~Pomarol, \emph{{Higgs as a holographic
  pseudoGoldstone boson}},
  \href{https://doi.org/10.1016/j.nuclphysb.2003.08.027}{\emph{Nucl. Phys. B}
  {\bfseries 671} (2003) 148--174},
  [\href{https://arxiv.org/abs/hep-ph/0306259}{{\ttfamily hep-ph/0306259}}].

\bibitem{Agashe:2004rs}
K.~Agashe, R.~Contino and A.~Pomarol, \emph{{The Minimal composite Higgs
  model}}, \href{https://doi.org/10.1016/j.nuclphysb.2005.04.035}{\emph{Nucl.
  Phys. B} {\bfseries 719} (2005) 165--187},
  [\href{https://arxiv.org/abs/hep-ph/0412089}{{\ttfamily hep-ph/0412089}}].

\bibitem{Barbieri:2023qpf}
R.~Barbieri and G.~Isidori, \emph{{Minimal flavour deconstruction}},
  \href{https://doi.org/10.1007/JHEP05(2024)033}{\emph{JHEP} {\bfseries 05}
  (2024) 033}, [\href{https://arxiv.org/abs/2312.14004}{{\ttfamily
  2312.14004}}].

\bibitem{Panico:2015jxa}
G.~Panico and A.~Wulzer, \emph{{The Composite Nambu-Goldstone Higgs}},
  vol.~913.
\newblock Springer, 2016,
  \href{https://doi.org/10.1007/978-3-319-22617-0}{10.1007/978-3-319-22617-0}.

\bibitem{Giudice:2007fh}
G.~F. Giudice, C.~Grojean, A.~Pomarol and R.~Rattazzi, \emph{{The
  Strongly-Interacting Light Higgs}},
  \href{https://doi.org/10.1088/1126-6708/2007/06/045}{\emph{JHEP} {\bfseries
  06} (2007) 045}, [\href{https://arxiv.org/abs/hep-ph/0703164}{{\ttfamily
  hep-ph/0703164}}].

\bibitem{Glioti:2024hye}
A.~Glioti, R.~Rattazzi, L.~Ricci and L.~Vecchi, \emph{{Exploring the Flavor
  Symmetry Landscape}},  \href{https://arxiv.org/abs/2402.09503}{{\ttfamily
  2402.09503}}.

\bibitem{Barbieri:2011ci}
R.~Barbieri, G.~Isidori, J.~Jones-Perez, P.~Lodone and D.~M. Straub,
  \emph{{$U(2)$ and Minimal Flavour Violation in Supersymmetry}},
  \href{https://doi.org/10.1140/epjc/s10052-011-1725-z}{\emph{Eur. Phys. J. C}
  {\bfseries 71} (2011) 1725},
  [\href{https://arxiv.org/abs/1105.2296}{{\ttfamily 1105.2296}}].

\bibitem{Barbieri:2012uh}
R.~Barbieri, D.~Buttazzo, F.~Sala and D.~M. Straub, \emph{{Flavour physics from
  an approximate $U(2)^3$ symmetry}},
  \href{https://doi.org/10.1007/JHEP07(2012)181}{\emph{JHEP} {\bfseries 07}
  (2012) 181}, [\href{https://arxiv.org/abs/1203.4218}{{\ttfamily 1203.4218}}].

\bibitem{Isidori:2012ts}
G.~Isidori and D.~M. Straub, \emph{{Minimal Flavour Violation and Beyond}},
  \href{https://doi.org/10.1140/epjc/s10052-012-2103-1}{\emph{Eur. Phys. J. C}
  {\bfseries 72} (2012) 2103},
  [\href{https://arxiv.org/abs/1202.0464}{{\ttfamily 1202.0464}}].

\bibitem{Redi:2012uj}
M.~Redi, \emph{{Composite MFV and Beyond}},
  \href{https://doi.org/10.1140/epjc/s10052-012-2030-1}{\emph{Eur. Phys. J. C}
  {\bfseries 72} (2012) 2030},
  [\href{https://arxiv.org/abs/1203.4220}{{\ttfamily 1203.4220}}].

\bibitem{Li:1981nk}
X.~Li and E.~Ma, \emph{{Gauge Model of Generation Nonuniversality}},
  \href{https://doi.org/10.1103/PhysRevLett.47.1788}{\emph{Phys. Rev. Lett.}
  {\bfseries 47} (1981) 1788}.

\bibitem{Dvali:2000ha}
G.~R. Dvali and M.~A. Shifman, \emph{{Families as neighbors in extra
  dimension}}, \href{https://doi.org/10.1016/S0370-2693(00)00083-6}{\emph{Phys.
  Lett. B} {\bfseries 475} (2000) 295--302},
  [\href{https://arxiv.org/abs/hep-ph/0001072}{{\ttfamily hep-ph/0001072}}].

\bibitem{Craig:2011yk}
N.~Craig, D.~Green and A.~Katz, \emph{{(De)Constructing a Natural and Flavorful
  Supersymmetric Standard Model}},
  \href{https://doi.org/10.1007/JHEP07(2011)045}{\emph{JHEP} {\bfseries 07}
  (2011) 045}, [\href{https://arxiv.org/abs/1103.3708}{{\ttfamily 1103.3708}}].

\bibitem{Panico:2016ull}
G.~Panico and A.~Pomarol, \emph{{Flavor hierarchies from dynamical scales}},
  \href{https://doi.org/10.1007/JHEP07(2016)097}{\emph{JHEP} {\bfseries 07}
  (2016) 097}, [\href{https://arxiv.org/abs/1603.06609}{{\ttfamily
  1603.06609}}].

\bibitem{Bordone:2017bld}
M.~Bordone, C.~Cornella, J.~Fuentes-Martin and G.~Isidori, \emph{{A three-site
  gauge model for flavor hierarchies and flavor anomalies}},
  \href{https://doi.org/10.1016/j.physletb.2018.02.011}{\emph{Phys. Lett. B}
  {\bfseries 779} (2018) 317--323},
  [\href{https://arxiv.org/abs/1712.01368}{{\ttfamily 1712.01368}}].

\bibitem{Greljo:2018tuh}
A.~Greljo and B.~A. Stefanek, \emph{{Third family quark\textendash{}lepton
  unification at the TeV scale}},
  \href{https://doi.org/10.1016/j.physletb.2018.05.033}{\emph{Phys. Lett. B}
  {\bfseries 782} (2018) 131--138},
  [\href{https://arxiv.org/abs/1802.04274}{{\ttfamily 1802.04274}}].

\bibitem{Fuentes-Martin:2020pww}
J.~Fuentes-Martin, G.~Isidori, J.~Pag\`es and B.~A. Stefanek, \emph{{Flavor
  non-universal Pati-Salam unification and neutrino masses}},
  \href{https://doi.org/10.1016/j.physletb.2021.136484}{\emph{Phys. Lett. B}
  {\bfseries 820} (2021) 136484},
  [\href{https://arxiv.org/abs/2012.10492}{{\ttfamily 2012.10492}}].

\bibitem{Fuentes-Martin:2020bnh}
J.~Fuentes-Mart\'\i{}n and P.~Stangl, \emph{{Third-family quark-lepton
  unification with a fundamental composite Higgs}},
  \href{https://doi.org/10.1016/j.physletb.2020.135953}{\emph{Phys. Lett. B}
  {\bfseries 811} (2020) 135953},
  [\href{https://arxiv.org/abs/2004.11376}{{\ttfamily 2004.11376}}].

\bibitem{Davighi:2022fer}
J.~Davighi and J.~Tooby-Smith, \emph{{Electroweak flavour unification}},
  \href{https://doi.org/10.1007/JHEP09(2022)193}{\emph{JHEP} {\bfseries 09}
  (2022) 193}, [\href{https://arxiv.org/abs/2201.07245}{{\ttfamily
  2201.07245}}].

\bibitem{Fuentes-Martin:2022xnb}
J.~Fuentes-Martin, G.~Isidori, J.~M. Lizana, N.~Selimovic and B.~A. Stefanek,
  \emph{{Flavor hierarchies, flavor anomalies, and Higgs mass from a warped
  extra dimension}},  \href{https://arxiv.org/abs/2203.01952}{{\ttfamily
  2203.01952}}.

\bibitem{FernandezNavarro:2022gst}
M.~Fern\'andez~Navarro and S.~F. King, \emph{{$B$-anomalies in a twin
  Pati-Salam theory of flavour}},
  \href{https://arxiv.org/abs/2209.00276}{{\ttfamily 2209.00276}}.

\bibitem{FernandezNavarro:2023rhv}
M.~Fern\'andez~Navarro and S.~F. King, \emph{{Tri-hypercharge: a separate
  gauged weak hypercharge for each fermion family as the origin of flavour}},
  \href{https://doi.org/10.1007/JHEP08(2023)020}{\emph{JHEP} {\bfseries 08}
  (2023) 020}, [\href{https://arxiv.org/abs/2305.07690}{{\ttfamily
  2305.07690}}].

\bibitem{Davighi:2022bqf}
J.~Davighi, G.~Isidori and M.~Pesut, \emph{{Electroweak-flavour and
  quark-lepton unification: a family non-universal path}},
  \href{https://arxiv.org/abs/2212.06163}{{\ttfamily 2212.06163}}.

\bibitem{Greljo:2024ovt}
A.~Greljo and G.~Isidori, \emph{{Neutrino Anarchy from Flavor Deconstruction}},
   \href{https://arxiv.org/abs/2406.01696}{{\ttfamily 2406.01696}}.

\bibitem{Capdevila:2024gki}
B.~Capdevila, A.~Crivellin, J.~M. Lizana and S.~Pokorski, \emph{{$SU(2)_L$
  deconstruction and flavour (non)-universality}},
  \href{https://arxiv.org/abs/2401.00848}{{\ttfamily 2401.00848}}.

\bibitem{Fuentes-Martin:2024fpx}
J.~Fuentes-Mart\'\i{}n and J.~M. Lizana, \emph{{Deconstructing flavor
  anomalously}}, \href{https://doi.org/10.1007/JHEP07(2024)117}{\emph{JHEP}
  {\bfseries 07} (2024) 117},
  [\href{https://arxiv.org/abs/2402.09507}{{\ttfamily 2402.09507}}].

\bibitem{Koppenburg:2023ndc}
P.~Koppenburg, \emph{{Flavour Physics at LHCb -- 50 years of the KM paradigm}},
   10, 2023, \href{https://arxiv.org/abs/2310.10504}{{\ttfamily 2310.10504}}.

\bibitem{Craig:2017cda}
N.~Craig, I.~Garcia~Garcia and D.~Sutherland, \emph{{Disassembling the
  Clockwork Mechanism}},
  \href{https://doi.org/10.1007/JHEP10(2017)018}{\emph{JHEP} {\bfseries 10}
  (2017) 018}, [\href{https://arxiv.org/abs/1704.07831}{{\ttfamily
  1704.07831}}].

\bibitem{Goursat1889}
E.~Goursat, \emph{Sur les substitutions orthogonales et les divisions
  régulières de l'espace}, {\emph{Annales scientifiques de l'École Normale
  Supérieure} {\bfseries 6} (1889) 9--102}.

\bibitem{bauer2015generalized}
K.~Bauer, D.~Sen and P.~Zvengrowski, \emph{{A Generalized Goursat Lemma}},
  \href{https://arxiv.org/abs/1109.0024}{{\ttfamily 1109.0024}}.

\bibitem{Allwicher:2020esa}
L.~Allwicher, G.~Isidori and A.~E. Thomsen, \emph{{Stability of the Higgs
  Sector in a Flavor-Inspired Multi-Scale Model}},
  \href{https://doi.org/10.1007/JHEP01(2021)191}{\emph{JHEP} {\bfseries 01}
  (2021) 191}, [\href{https://arxiv.org/abs/2011.01946}{{\ttfamily
  2011.01946}}].

\bibitem{Davighi:2023evx}
J.~Davighi and B.~A. Stefanek, \emph{{Deconstructed hypercharge: a natural
  model of flavour}},
  \href{https://doi.org/10.1007/JHEP11(2023)100}{\emph{JHEP} {\bfseries 11}
  (2023) 100}, [\href{https://arxiv.org/abs/2305.16280}{{\ttfamily
  2305.16280}}].

\bibitem{Davighi:2023xqn}
J.~Davighi, A.~Gosnay, D.~J. Miller and S.~Renner, \emph{{Phenomenology of a
  Deconstructed Electroweak Force}},
  \href{https://arxiv.org/abs/2312.13346}{{\ttfamily 2312.13346}}.

\bibitem{Chung:2021ekz}
Y.~Chung, \emph{{Flavorful composite Higgs model: Connecting the B anomalies
  with the hierarchy problem}},
  \href{https://doi.org/10.1103/PhysRevD.104.115027}{\emph{Phys. Rev. D}
  {\bfseries 104} (2021) 115027},
  [\href{https://arxiv.org/abs/2108.08511}{{\ttfamily 2108.08511}}].

\bibitem{Chung:2021fpc}
Y.~Chung, \emph{{Composite flavon-Higgs models}},
  \href{https://doi.org/10.1103/PhysRevD.104.095011}{\emph{Phys. Rev. D}
  {\bfseries 104} (2021) 095011},
  [\href{https://arxiv.org/abs/2104.11719}{{\ttfamily 2104.11719}}].

\bibitem{Bally:2022naz}
A.~Bally, Y.~Chung and F.~Goertz, \emph{{Hierarchy problem and the top Yukawa
  coupling: An alternative to top partner solutions}},
  \href{https://doi.org/10.1103/PhysRevD.108.055008}{\emph{Phys. Rev. D}
  {\bfseries 108} (2023) 055008},
  [\href{https://arxiv.org/abs/2211.17254}{{\ttfamily 2211.17254}}].

\bibitem{Chung:2023gcm}
Y.~Chung and F.~Goertz, \emph{{Third-generation-philic Hidden Naturalness}},
  \href{https://arxiv.org/abs/2311.17169}{{\ttfamily 2311.17169}}.

\bibitem{Chung:2023iwj}
Y.~Chung, \emph{{Naturalness-motivated composite Higgs model for generating the
  top Yukawa coupling}},
  \href{https://doi.org/10.1103/PhysRevD.109.095021}{\emph{Phys. Rev. D}
  {\bfseries 109} (2024) 095021},
  [\href{https://arxiv.org/abs/2309.00072}{{\ttfamily 2309.00072}}].

\bibitem{Gripaios:2009pe}
B.~Gripaios, A.~Pomarol, F.~Riva and J.~Serra, \emph{{Beyond the Minimal
  Composite Higgs Model}},
  \href{https://doi.org/10.1088/1126-6708/2009/04/070}{\emph{JHEP} {\bfseries
  04} (2009) 070}, [\href{https://arxiv.org/abs/0902.1483}{{\ttfamily
  0902.1483}}].

\bibitem{Setford:2017csx}
J.~Setford, \emph{{Composite Higgs models in disguise}},
  \href{https://doi.org/10.1007/JHEP01(2018)092}{\emph{JHEP} {\bfseries 01}
  (2018) 092}, [\href{https://arxiv.org/abs/1710.11206}{{\ttfamily
  1710.11206}}].

\bibitem{Davighi:2018xwn}
J.~Davighi and B.~Gripaios, \emph{{Topological terms in Composite Higgs
  Models}}, \href{https://doi.org/10.1007/JHEP11(2018)169}{\emph{JHEP}
  {\bfseries 11} (2018) 169},
  [\href{https://arxiv.org/abs/1808.04154}{{\ttfamily 1808.04154}}].

\bibitem{Gripaios:2014pqa}
B.~Gripaios, T.~M\"uller, M.~A. Parker and D.~Sutherland, \emph{{Search
  Strategies for Top Partners in Composite Higgs models}},
  \href{https://doi.org/10.1007/JHEP08(2014)171}{\emph{JHEP} {\bfseries 08}
  (2014) 171}, [\href{https://arxiv.org/abs/1406.5957}{{\ttfamily 1406.5957}}].

\bibitem{Agashe:2006at}
K.~Agashe, R.~Contino, L.~Da~Rold and A.~Pomarol, \emph{{A Custodial symmetry
  for $Zb \bar b$}},
  \href{https://doi.org/10.1016/j.physletb.2006.08.005}{\emph{Phys. Lett. B}
  {\bfseries 641} (2006) 62--66},
  [\href{https://arxiv.org/abs/hep-ph/0605341}{{\ttfamily hep-ph/0605341}}].

\bibitem{Farina:2013mla}
M.~Farina, D.~Pappadopulo and A.~Strumia, \emph{{A modified naturalness
  principle and its experimental tests}},
  \href{https://doi.org/10.1007/JHEP08(2013)022}{\emph{JHEP} {\bfseries 08}
  (2013) 022}, [\href{https://arxiv.org/abs/1303.7244}{{\ttfamily 1303.7244}}].

\bibitem{Callan:1969sn}
C.~G. Callan, Jr., S.~R. Coleman, J.~Wess and B.~Zumino, \emph{{Structure of
  phenomenological Lagrangians. 2.}},
  \href{https://doi.org/10.1103/PhysRev.177.2247}{\emph{Phys. Rev.} {\bfseries
  177} (1969) 2247--2250}.

\bibitem{Gripaios:2015qya}
B.~Gripaios, \emph{{Lectures on Effective Field Theory}},
  \href{https://arxiv.org/abs/1506.05039}{{\ttfamily 1506.05039}}.

\bibitem{Durieux:2021riy}
G.~Durieux, M.~McCullough and E.~Salvioni, \emph{{Gegenbauer Goldstones}},
  \href{https://doi.org/10.1007/JHEP01(2022)076}{\emph{JHEP} {\bfseries 01}
  (2022) 076}, [\href{https://arxiv.org/abs/2110.06941}{{\ttfamily
  2110.06941}}].

\bibitem{Durieux:2022sgm}
G.~Durieux, M.~McCullough and E.~Salvioni, \emph{{Gegenbauer\textquoteright{}s
  Twin}}, \href{https://doi.org/10.1007/JHEP05(2022)140}{\emph{JHEP} {\bfseries
  05} (2022) 140}, [\href{https://arxiv.org/abs/2202.01228}{{\ttfamily
  2202.01228}}].

\bibitem{Panico:2012uw}
G.~Panico, M.~Redi, A.~Tesi and A.~Wulzer, \emph{{On the Tuning and the Mass of
  the Composite Higgs}},
  \href{https://doi.org/10.1007/JHEP03(2013)051}{\emph{JHEP} {\bfseries 03}
  (2013) 051}, [\href{https://arxiv.org/abs/1210.7114}{{\ttfamily 1210.7114}}].

\bibitem{Csaki:2017cep}
C.~Csaki, T.~Ma and J.~Shu, \emph{{Maximally Symmetric Composite Higgs
  Models}}, \href{https://doi.org/10.1103/PhysRevLett.119.131803}{\emph{Phys.
  Rev. Lett.} {\bfseries 119} (2017) 131803},
  [\href{https://arxiv.org/abs/1702.00405}{{\ttfamily 1702.00405}}].

\bibitem{Contino:2010rs}
R.~Contino, \emph{{The Higgs as a Composite Nambu-Goldstone Boson}},  in
  \emph{{Theoretical Advanced Study Institute in Elementary Particle Physics}:
  {Physics of the Large and the Small}}, pp.~235--306, 2011,
  \href{https://arxiv.org/abs/1005.4269}{{\ttfamily 1005.4269}},
  \href{https://doi.org/10.1142/9789814327183_0005}{DOI}.

\bibitem{ATLAS:2020qdt}
{\scshape ATLAS} collaboration, \emph{{A combination of measurements of Higgs
  boson production and decay using up to $139$ fb$^{-1}$ of proton--proton
  collision data at $\sqrt{s}=$ 13 TeV collected with the ATLAS experiment}}, .

\bibitem{CMS:2020gsy}
{\scshape CMS} collaboration, \emph{{Combined Higgs boson production and decay
  measurements with up to 137 fb$^{-1}$ of proton-proton collision data at
  $\sqrt s$ = 13 TeV}}, .

\bibitem{ATLAS:2024gyc}
{\scshape ATLAS} collaboration, G.~Aad et~al., \emph{{Search for
  pair-production of vector-like quarks in lepton+jets final states containing
  at least one b-tagged jet using the Run 2 data from the ATLAS experiment}},
  \href{https://doi.org/10.1016/j.physletb.2024.138743}{\emph{Phys. Lett. B}
  {\bfseries 854} (2024) 138743},
  [\href{https://arxiv.org/abs/2401.17165}{{\ttfamily 2401.17165}}].

\bibitem{CMS:2022fck}
{\scshape CMS} collaboration, \emph{{Search for pair production of vector-like
  quarks in leptonic final states in proton-proton collisions at $\sqrt{s}$ =
  13 TeV}},  \href{https://arxiv.org/abs/2209.07327}{{\ttfamily 2209.07327}}.

\bibitem{ATLAS:2023bfh}
{\scshape ATLAS} collaboration, G.~Aad et~al., \emph{{Search for singly
  produced vectorlike top partners in multilepton final states with 139\,\,fb-1
  of pp collision data at s=13\,\,TeV with the ATLAS detector}},
  \href{https://doi.org/10.1103/PhysRevD.109.112012}{\emph{Phys. Rev. D}
  {\bfseries 109} (2024) 112012},
  [\href{https://arxiv.org/abs/2307.07584}{{\ttfamily 2307.07584}}].

\bibitem{CMS:2024bni}
{\scshape CMS} collaboration, A.~Hayrapetyan et~al., \emph{{Review of searches
  for vector-like quarks, vector-like leptons, and heavy neutral leptons in
  proton-proton collisions at $\sqrt{s}$ = 13 TeV at the CMS experiment}},
  \href{https://arxiv.org/abs/2405.17605}{{\ttfamily 2405.17605}}.

\bibitem{ATLAS:2024qvg}
{\scshape ATLAS} collaboration, G.~Aad et~al., \emph{{Combination of searches
  for heavy spin-1 resonances using 139 ${\rm fb}^{-1}$ of proton-proton
  collision data at $\sqrt{s}$=13 TeV with the ATLAS detector}},
  \href{https://doi.org/10.1007/JHEP04(2024)118}{\emph{JHEP} {\bfseries 04}
  (2024) 118}, [\href{https://arxiv.org/abs/2402.10607}{{\ttfamily
  2402.10607}}].

\bibitem{CMS:2023gte}
{\scshape CMS} collaboration, A.~Hayrapetyan et~al., \emph{{Search for W'
  bosons decaying to a top and a bottom quark in leptonic final states in
  proton-proton collisions at $ \sqrt{s} $ = 13 TeV}},
  \href{https://doi.org/10.1007/JHEP05(2024)046}{\emph{JHEP} {\bfseries 05}
  (2024) 046}, [\href{https://arxiv.org/abs/2310.19893}{{\ttfamily
  2310.19893}}].

\bibitem{Cornella:2021sby}
C.~Cornella, D.~A. Faroughy, J.~Fuentes-Martin, G.~Isidori and M.~Neubert,
  \emph{{Reading the footprints of the B-meson flavor anomalies}},
  \href{https://doi.org/10.1007/JHEP08(2021)050}{\emph{JHEP} {\bfseries 08}
  (2021) 050}, [\href{https://arxiv.org/abs/2103.16558}{{\ttfamily
  2103.16558}}].

\bibitem{deBlas:2017xtg}
J.~de~Blas, J.~C. Criado, M.~Perez-Victoria and J.~Santiago, \emph{{Effective
  description of general extensions of the Standard Model: the complete
  tree-level dictionary}},
  \href{https://doi.org/10.1007/JHEP03(2018)109}{\emph{JHEP} {\bfseries 03}
  (2018) 109}, [\href{https://arxiv.org/abs/1711.10391}{{\ttfamily
  1711.10391}}].

\bibitem{Bjorn:2016zlr}
M.~Bj\o{}rn and M.~Trott, \emph{{Interpreting $W$ mass measurements in the
  SMEFT}}, \href{https://doi.org/10.1016/j.physletb.2016.10.003}{\emph{Phys.
  Lett. B} {\bfseries 762} (2016) 426--431},
  [\href{https://arxiv.org/abs/1606.06502}{{\ttfamily 1606.06502}}].

\bibitem{LHC-TeVMWWorkingGroup:2023zkn}
{\scshape LHC-TeV~MW~Working~Group} collaboration, S.~Amoroso et~al.,
  \emph{{Compatibility and combination of world W-boson mass measurements}},
  \href{https://doi.org/10.1140/epjc/s10052-024-12532-z}{\emph{Eur. Phys. J. C}
  {\bfseries 84} (2024) 451},
  [\href{https://arxiv.org/abs/2308.09417}{{\ttfamily 2308.09417}}].

\bibitem{ATLAS:2020zms}
{\scshape ATLAS} collaboration, G.~Aad et~al., \emph{{Search for heavy Higgs
  bosons decaying into two tau leptons with the ATLAS detector using $pp$
  collisions at $\sqrt{s}=13$ TeV}},
  \href{https://doi.org/10.1103/PhysRevLett.125.051801}{\emph{Phys. Rev. Lett.}
  {\bfseries 125} (2020) 051801},
  [\href{https://arxiv.org/abs/2002.12223}{{\ttfamily 2002.12223}}].

\bibitem{CMS:2021ctt}
{\scshape CMS} collaboration, A.~M. Sirunyan et~al., \emph{{Search for resonant
  and nonresonant new phenomena in high-mass dilepton final states at $
  \sqrt{s} $ = 13 TeV}},
  \href{https://doi.org/10.1007/JHEP07(2021)208}{\emph{JHEP} {\bfseries 07}
  (2021) 208}, [\href{https://arxiv.org/abs/2103.02708}{{\ttfamily
  2103.02708}}].

\bibitem{Greljo:2017vvb}
A.~Greljo and D.~Marzocca, \emph{{High-$p_T$ dilepton tails and flavor
  physics}}, \href{https://doi.org/10.1140/epjc/s10052-017-5119-8}{\emph{Eur.
  Phys. J. C} {\bfseries 77} (2017) 548},
  [\href{https://arxiv.org/abs/1704.09015}{{\ttfamily 1704.09015}}].

\bibitem{Allwicher:2022gkm}
L.~Allwicher, D.~A. Faroughy, F.~Jaffredo, O.~Sumensari and F.~Wilsch,
  \emph{{Drell-Yan tails beyond the Standard Model}},
  \href{https://doi.org/10.1007/JHEP03(2023)064}{\emph{JHEP} {\bfseries 03}
  (2023) 064}, [\href{https://arxiv.org/abs/2207.10714}{{\ttfamily
  2207.10714}}].

\bibitem{Allwicher:2022mcg}
L.~Allwicher, D.~A. Faroughy, F.~Jaffredo, O.~Sumensari and F.~Wilsch,
  \emph{{HighPT: A tool for~ high-$p_T$ Drell-Yan tails beyond the standard
  model}}, \href{https://doi.org/10.1016/j.cpc.2023.108749}{\emph{Comput. Phys.
  Commun.} {\bfseries 289} (2023) 108749},
  [\href{https://arxiv.org/abs/2207.10756}{{\ttfamily 2207.10756}}].

\bibitem{Stefanek:2024kds}
B.~A. Stefanek, \emph{{Non-universal probes of composite Higgs models: new
  bounds and prospects for FCC-ee}},
  \href{https://doi.org/10.1007/JHEP09(2024)103}{\emph{JHEP} {\bfseries 09}
  (2024) 103}, [\href{https://arxiv.org/abs/2407.09593}{{\ttfamily
  2407.09593}}].

\end{thebibliography}\endgroup
%%%%%%%%%%%%%%%%%%%%%%%%%%%%%%%%%%%%%%%%%%%%%%%%%%%%%%%%%

\end{document}